\documentclass[conference]{IEEEtran}
\IEEEoverridecommandlockouts

\usepackage{hyperref}
\usepackage{url}
\usepackage{color}
\usepackage{amssymb}
\usepackage{amsmath}
\usepackage{amsthm}
\usepackage{listings}
\usepackage{graphicx}
\usepackage{subfigure}
\usepackage{comment}
\usepackage{soul}
\usepackage[noend]{distribalgo}
\usepackage{wrapfig}

\usepackage[colorinlistoftodos, textwidth=4cm, shadow, disable]{todonotes}

\usepackage{tikz}

\usepackage{comment}

\usepackage{algorithm}
\usepackage{algorithmic}
\usepackage{bbm}

\usepackage{multicol}
\usepackage{multirow}

\usepackage{fixme}
\fxsetup{
    status=draft,
    author={TODO},
    layout=inline, 
    theme=color
}

\usepackage{amsthm}

\newcommand{\FfDL}{FfDL}

\newcommand{\JSA}{JSA}

\newcommand{\myalgospace}{\vspace{1mm}}
\newcommand{\set}[1]{\ensuremath{\{ #1 \}}}

\newcommand{\executing}{\textsc{Executing}}
\newcommand{\arrived}{\textsc{Arrived}}
\newcommand{\finished}{\textsc{Finished}}

\newcommand{\trial}{\textsc{Trial}}

\newcommand{\process}[1]{\textsc{Process}(#1)}

\newcommand{\arrival}[1]{\textsc{Arrival}(\ensuremath{#1})}
\newcommand{\departure}[1]{\textsc{Departure}(\ensuremath{#1})}

\newcommand{\job}[1][]{\ensuremath{job}_{#1}}
\newcommand{\enqueue}[2]{\ensuremath{\textsc{Enqueue}(#1, #2)}}
\newcommand{\remove}[2]{\ensuremath{\textsc{Remove}(#1, #2)}}

\definecolor{mygreen}{rgb}{0,0.6,0}
\definecolor{mygray}{rgb}{0.5,0.5,0.5}
\definecolor{mymauve}{rgb}{0.58,0,0.82}

\renewcommand{\fxfatal}[1]{}
\renewcommand{\fxwarning}[1]{}

\begin{document}

%


\title{Effective Elastic Scaling of Deep Learning Workloads}



\author{\IEEEauthorblockN{Vaibhav Saxena}
\textit{IBM Research, India}
\and
\IEEEauthorblockN{K. R. Jayaram}
\textit{IBM Research, USA}
\and
\IEEEauthorblockN{Saurav Basu}
\textit{Microsoft, India}
\and
\IEEEauthorblockN{Yogish Sabharwal}
\textit{IBM Research, India}
\and
\IEEEauthorblockN{Ashish Verma}
\textit{IBM Research, USA}}


%
\maketitle

\thispagestyle{plain}
\pagestyle{plain}

%
\begin{abstract}
The increased use of deep learning (DL) in academia, government and industry has, in turn, led to the popularity of on-premise and cloud-hosted deep learning platforms, whose goals are to enable organizations utilize expensive resources effectively, and to share said resources among multiple teams in a fair and effective manner. 

In this paper, we examine the elastic scaling of Deep Learning (DL) jobs over large-scale training platforms 
and propose a novel resource allocation strategy for DL training jobs, resulting in improved  
job run time performance as well as increased  cluster utilization. We begin by analyzing DL workloads and exploit the fact that DL jobs can be run with a range of batch 
sizes without affecting their final accuracy. We formulate an optimization problem that 
explores a dynamic batch size allocation to individual DL jobs based on their scaling efficiency, when running on multiple nodes. 
We design a fast dynamic programming based optimizer to solve this problem in real-time
to determine jobs that can be scaled up/down, and use this optimizer in 
an autoscaler to dynamically change the allocated resources and batch sizes of
individual DL jobs. 

We demonstrate empirically that our elastic scaling algorithm can complete up to 
$\approx 2 \times$ as many jobs as compared to a strong baseline algorithm that also scales the number of GPUs but does not change the batch size. We also demonstrate that the average completion time with our algorithm is up to
$\approx 10 \times$ faster than that of the baseline.

\end{abstract}


\section{Introduction}~\label{Sec:intro}

Deep Learning (DL) models can achieve state-of-the-art accuracy, sometimes exceeding human-level performance~\cite{lecun:15} in image, video, speech and text recognition with applications to driver-less cars, voice control in consumer devices, video search, social networks, etc. The rise of easy-to-use, open-source
DL frameworks like TensorFlow~\cite{tensorflow}, PyTorch~\cite{pytorch} and
Caffe~\cite{caffe2} have further enhanced the popularity of DL.  The increased use of DL has, in turn, led to the popularity of on-premise and cloud-hosted DL platforms, whose goals are to enable organizations utilize expensive resources effectively, and to share said resources among multiple teams.

Many existing elastic scaling 
techniques~\cite{abhishek-mapred-elasticity, chase-elasticstorage, google-wavelets, anshul-autoscale, bauer-chameleon, herbst-tompecs-metric}
are hard to apply to 
DL platforms/clusters because they focus on \emph{cluster elasticity} -- scaling the size of the cluster with VMs 
in response to workload variations. 
DL workloads and their system (hardware/software) stack present unique challenges for cluster elasticity.
DL training is generally performed
on parallel systems using high performance GPU accelerators, due to 
their efficiency in matrix computations and convolutions, and high (GPU) memory
bandwidth. 
Also, many DL frameworks and workloads
benefit from and are optimized for high speed interconnection networks (like NVLink~\cite{nvlink},
Infiniband~\cite{infiniband}, RDMA and 100G Ethernet). To make effective use of 
such interconnects, DL workloads need \emph{locality aware scheduling}~\cite{fiddle-atc}, i.e., 
learning processes in a distributed DL job have to be located as close to
each other as possible. So, unless the entire datacenter has the same
hardware (GPUs) and interconnect,  
it becomes hard to \emph{quickly} scale a DL cluster. Often, one cannot scale the cluster by acquiring VMs,
even GPU enabled VMs, because they can be provisioned anywhere in the datacenter. Cluster scaling
with support for locality aware placement and scheduling requires manual intervention, installation and interconnect configuration taking
several hours~\cite{projectphilly, fiddle-atc}. Second, there are often cost-related reasons why DL cluster scaling is undesirable.

Hence, elasticity is vital to optimize resource utilization and improve user
experience of DL training clusters,
which, like other clusters, experience wide variations in workloads.
For example, sometimes the cluster may become underutilized when the number of jobs is very low leading to idle resources. Similarly, the cluster may become oversubscribed and some jobs are unable to run due to shortage of resources.
Hence, in this paper, we consider \emph{job elasticity}~\cite{spark, hadoopmr}, where jobs
themselves are scaled to fit a fixed size cluster. We leverage a key
characteristic of DL jobs -- \emph{batch size} -- for job elasticity.

Batch size is the number of samples used to determine the gradient of the parameters of the model at each update step of the gradient descent algorithm~\cite{bottou}. A very small batch size may result in noisy gradients which takes large number of iterations to converge, while using very large batch sizes has shown to result in poor generalization~\cite{keskar}. Just as an example for training neural networks on image datasets, researchers have typically used batch-sizes in the range from 32 to 8192~\cite{goyal}. Although recently there is evidence to be able use much larger batch-sizes (upto 32K) without affecting the accuracy significantly~\cite{you}. Furthermore, researchers have shown that instead of keeping a fixed batch-size over the training period, the batch-size can be changed dynamically without affecting the model accuracy~\cite{smith-1,adabatch,balles}. We also performed experiments demonstrating this behaviour which are discussed in Section~\ref{sec:minmax}.


Batch-size has a direct implication on using the compute resources in the cluster and hence the time to complete the training process. In other words, by increasing the batch-size a higher number of GPUs can be leveraged to finish the training faster and vice-versa.

The goal and main contribution of this paper is to \emph{use permissible 
batch size ranges for performing efficient resource allocation and
elastic scaling} of DL jobs. To this end, this paper makes the following 
technical contributions:

\begin{enumerate}
    \item A \emph{job scalability analyzer} to determine the scaling 
    characteristics of DL jobs and their runtime with respect to the 
    number of GPUs allotted and batch sizes used (Section~\ref{sec:runtimeestimation}).
    
    \item A \emph{fast, dynamic programming based optimizer} that uses the job scalability analyzer to allocate 
    optimal resources and batch sizes to DL jobs \emph{in real time} to maximize cluster throughput (Section~\ref{sec:optimizer}).
    
    \item An \emph{autoscaler} that uses the optimizer, cluster metrics and user input
    to elastically scale jobs using checkpoint-resume (Section~\ref{sec:autoscaler}). We 
    demonstrate that the design and implementation of the autoscaler is independent 
    of the middleware used to manage DL training jobs.

    \item A detailed empirical evaluation (Section~\ref{sec:eval}) of the efficacy of our elastic scaling algorithm 
    demonstrating that it is able to complete up to $\approx 2\times$ the number of jobs in comparison to the
    baseline algorithm that does not consider or vary batch size. We also demonstrate that when queueing of jobs due 
    to resource scarcity is not possible/desired, our elastic scaling algorithm drops up to $\approx 3 \times$ fewer
    jobs, and that the average job completion time with our algorithm is up to $\approx 10 \times$ better than that of the 
    baseline when job queueing is enabled.
\end{enumerate}

\section{Background \& Architecture}

This is a paper on effective elastic scaling mechanisms for DL training 
workloads. We present our resource allocation and elastic scaling techniques in the context of DL training platform.

\subsection{DL Training Platforms (DLP)}

A DLP is a multi-user, multi-tenant middleware for deep learning training.
Architecturally, it resides above a cluster manager like Kubernetes~\cite{kubernetes}
or Mesos~\cite{mesos} and uses the cluster manager for deployment, placement, and failure
detection of training jobs. The goals of a DL platform are (i) to enable data scientists to focus only on their training 
application, and not worry about any setup, security and failure handling, (ii)
to enable organizations to share expensive DL hardware, and (iii) to serve as a 
backbone for commercial DL as-a-Service offerings.

DL platforms are responsible for training jobs from start to finish, and should 
be highly available, scalable and efficient. For this paper, we use 
\FfDL (Framework for Deep Learning)~\cite{ffdl, ffdl-exp} which is an 
\emph{open-source} DL platform.
A detailed description of design and implementation
of \FfDL\ is beyond the scope of this paper and we refer the reader to \cite{ffdl, ffdl-exp}.
Our autoscaler only requires the DLP to support
creation, deletion, halting (with a checkpoint) and 
resumption of DL training jobs; many DLPs already do this~\cite{fiddle-atc}. The actual job scaling happens through halt/resume -- halting
the job that is currently running and resuming the same from its last checkpoint.
\FfDL\ employs Kubernetes~\cite{kubernetes} (K8S)
for container orchestration and cluster management.
This enables \FfDL\ to create replicated 
learners (for distributed training) and is well suited for DL frameworks like Horovod and
distributed Tensorflow.

\begin{figure}[htb]
\centering
\includegraphics[width=0.7\columnwidth,keepaspectratio=true]{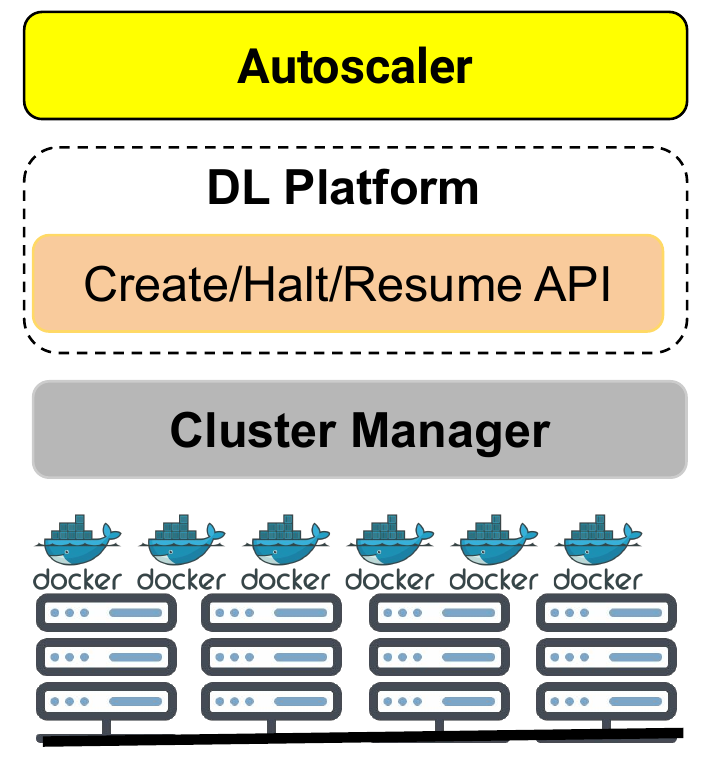}
\caption{High-level architecture of our techniques}~\label{fig:dlparch}
\end{figure}


\subsection{DL Training Jobs}

Data scientists typically write DL training jobs using a DL framework
like Caffe~\cite{caffe}, PyTorch~\cite{pytorch}, TensorFlow~\cite{tensorflow},
Horovod~\cite{horovod}, etc., which internally use different communication libraries (MPI/OpenMP/Gloo).  
To abstract away framework/communication specific details, and simplify job management,
DL platforms, including \FfDL, often use Docker containers~\cite{docker}.
\FfDL\ takes user code, and 
automatically instantiates  Docker containers (from its
library of Docker images) containing 
everything necessary to run the training job.

A DL training job \emph{typically} consists of a set of 
learning processes (``learners''),
with each learner running in a Docker container using one or more GPUs.
Each learner container consists of a framework Docker image (Caffe, TensorFlow, etc.)
instantiated with user code, job parameters, locations where training data can be 
accessed and where checkpoints/logs/results have to stored, and credentials (certificates, keys, etc.)
used for communication, data access and storing results. 

Each learner comprises of a copy of the neural network model that is being trained. The model has a set of weights that are \emph{learned} through time, and an associated model architecture which defines the rules to update the weights. The values of weight updates at an iteration is called the gradient. Each learner works on a subset of the entire data at every iteration and prepares its copy of the gradient, which is then combined to form a single gradient through a learner-to-learner communication call such as AllReduce at every iteration. Each worker updates its copy of the model weights with the combined gradient at every iteration. 

Communication and synchronization methods internal to frameworks 
(i.e., MPI, OpenMP, use of specific parameter servers, etc.) are preserved 
while the job is executed in the DL platform and are part of the framework Docker image.

Job parameters, including the source of training data,
credentials to access training data, framework, number of learners, 
location where results and logs should be stored, learning rate,
etc., are specified by the user/data scientist using a manifest file. 


\subsection{Min- and Max- Batch Size}~\label{sec:minmax}

An important characteristic of deep learning jobs is the batch size. The batch size defines the number of samples that are processed by the network in order to collectively determine the gradient for performing the update step in the gradient descent algorithm. If the batch size is too large, then the training runs very slowly whereas if the batch size is too small, the accuracy often degrades (references). Batch size is a very important parameter when scaling deep learning jobs. When the number of nodes is increased, users often increase the batch size linearly as this leads to improved resource utilization. As mentioned earlier, prior work has not dealt with taking into account or modifying the job parameters while scheduling/elastically scaling the jobs.

\begin{figure}[htb]
\centering
\includegraphics[width=\columnwidth]{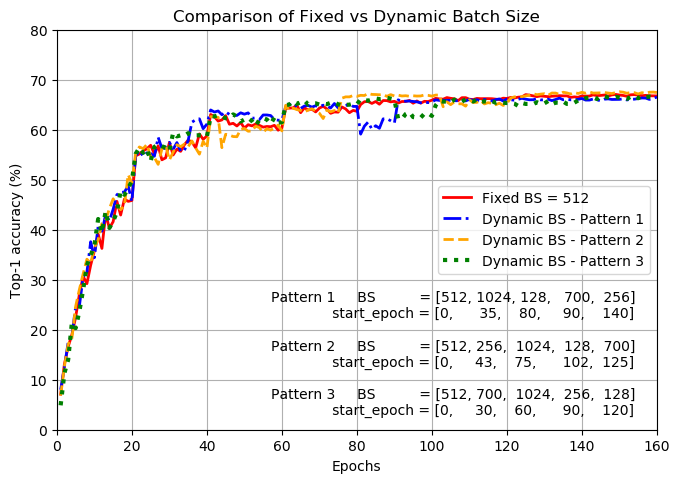}
\caption{Similarity of final accuracy in DL jobs when the batch size (BS) varies within an allowable range. Three BS variation patterns are used. \emph{start\_epoch} lists the starting epoch for the corresponding BS change.} 
\label{fig:upNdown}
\end{figure}

As mentioned in the Introduction, a peculiar characteristic of a DL job is that the final accuracy of the model remains similar if the training is done within a certain range of batchsizes, called the optimal batchsize range. This has been previously noted in some recent research \cite{smith-1, goyal, keskar}.

We have experimented extensively in continuance to the hypothesis in \cite{smith-1, goyal, keskar} to verify that both increasing and decreasing the batchsize within the tolerable range does not affect the final convergence. One such example is shown in Figure \ref{fig:upNdown}. The model is VGG11 \cite{vgg} which is trained on the dataset CIFAR100 \cite{cifar} for 160 epochs. The curve in red (Fixed BS) is a usual training with a fixed batchsize of 512 images throughout the training. The other three curves (Dynamic BS) show a training where the batchsize has been increased or decreased at random epochs, but within a range of 128-1024 images. The comparison of the curves show that both fixed BS and dynamic (varying) BS lead to similar final accuracy ($\pm$ 0.5\%).


\subsection{Assumptions}

{\bf Focus on GPU Allocation: }Most DL platforms (\FfDL\ included)
recommend using a specific number of CPU cores and specific amount of RAM corresponding to each GPU type~\cite{ffdl, ffdl-exp},
in order to maximize performance.
For example, the DL platform may recommend using 4 CPU cores and 8GB RAM when 1 K80 is used by the DL
job, and 6 CPU cores and 12 GB RAM when 1 V100 is used. It has been observed~\cite{ffdl-exp} that data scientists do not
deviate from these recommendations. So, it is typically sufficient to first solve GPU allocation and use the platform's 
guidelines on proportionally scaling CPUs and RAM.


\noindent{\bf Synchronous DL: }Our results are presented for the typical case of 
synchronous distributed DL, where all the learners work on the 
exact same copy of the model weights. This is a much more common paradigm than the sometimes 
used asynchronous distributed DL paradigm, where the model updates to separate learners happen independently. 
Synchronous DL not only leads to generally better convergence, but the results are easy 
to reproduce.

\noindent{\bf Homogeneous Distributed Jobs: } We also assume that distributed DL jobs are homogeneous, i.e.,
each learner in a job uses the same type of GPU and 
jobs do not contain learners using a mix of say K80 and P100 GPUs.
This is quite typical in DL training; given vast differences in compute 
power between different GPU types, distributed jobs are mainly run in homogeneous fashion for effective data parallelism. Second, while our techniques do not require an entire
datacenter to be homogeneous with respect to the types of GPUs;
we assume that the datacenter is divided into clusters
of homogeneous GPUs, and each autoscaler only manages homogeneous GPUs.

\section{DL-aware Elastic Scaling}

\subsection{Overview and Design Principles}~\label{sec:designoverview} 

\begin{figure}[htb]
\includegraphics[width=\columnwidth, keepaspectratio=true]{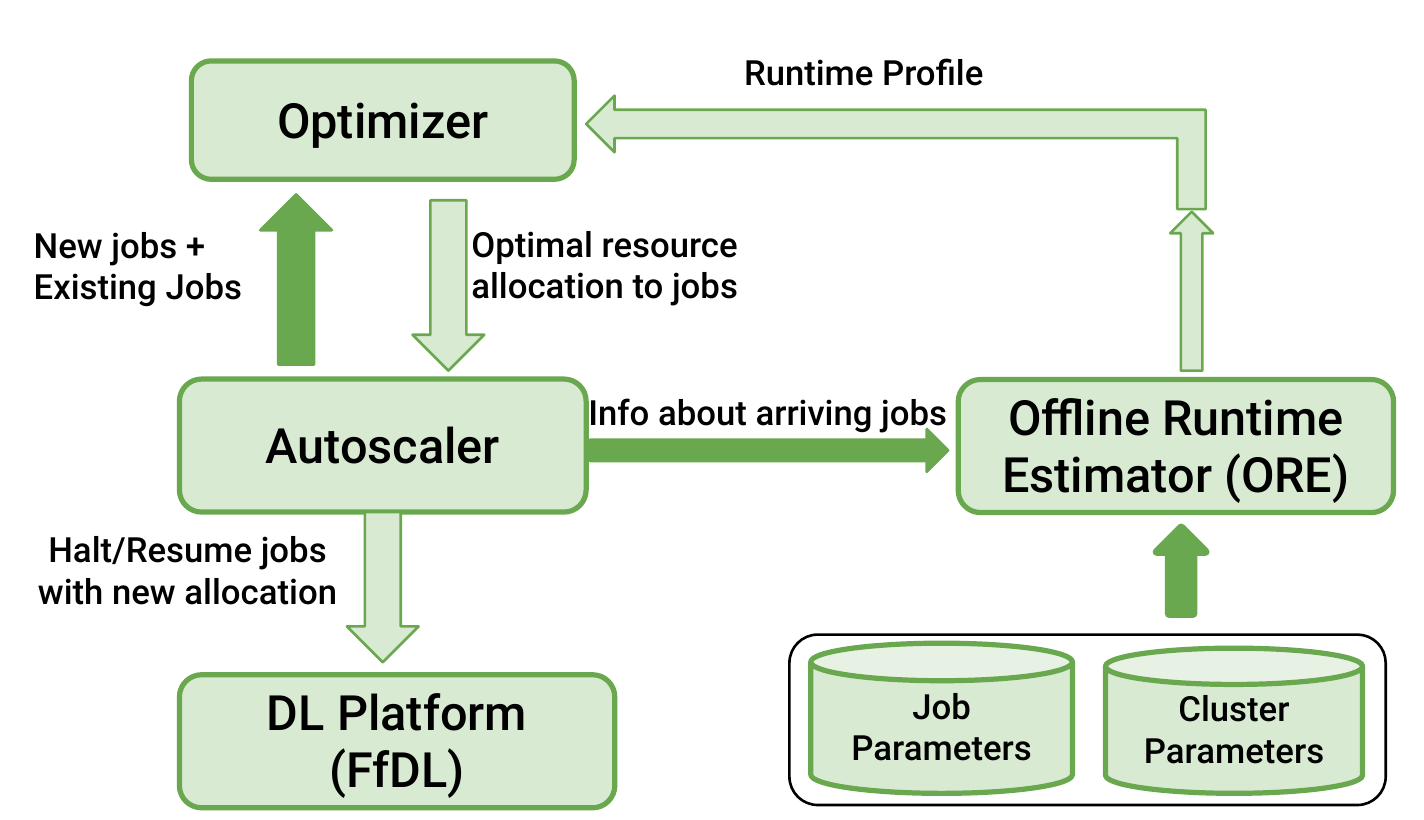}
\caption{Overview of our DL-aware elastic mechanism}~\label{fig:overview}
\end{figure}

Our mechanism is composed of three core modules 
- (i) a \emph{job scalability analyzer} (JSA), (ii) an \textit{optimizer} and
(iii) an \textit{autoscaler}. The optimizer's goal is to allocate GPUs
to the current set of jobs in the cluster in an (optimal) manner that 
maximizes cluster utilization and throughput (c.f. Section~\ref{sec:optimizer} for the precise 
formulation of throughput; informally it captures overall progress). The role of the JSA 
is to determine the scaling characteristics
of the jobs in order to estimate their runtime so that the optimizer can make
optimal scheduling decisions. The goals of the autoscaler are to (i) receive and buffer
incoming jobs, (ii) periodically consult the optimizer to determine the set
of jobs that can be allocated resources, and the number of GPUs to allot to
each of the jobs in the set, and (iii) interact with the DL platform to scale 
jobs up or down.

Our mechanism is designed to be agnostic to the internals of specific DL platforms.
The autoscaler, in conjunction with the runtime estimator and optimizer
performs \emph{resource allocation}; placement of jobs on nodes is performed
by the DL platform and the cluster manager inside the DL platform.

In contrast to existing DL cloud scheduling systems, our system also takes 
the minimum and the maximum batchsize ($b_{min}$ and $b_{\mathit{max}}$ respectively),
as additional user specified parameters as part of the input job specification.
In addition, it allows jobs to be scheduled up to a maximum number of $k_{\mathit{max}}$ GPUs.
\todo[inline, color=green]{Is it okay to vary batch size with impunity? }
\todo[inline, color=green]{How does the data scientist know which batch size ranges to choose?}
\todo[inline, color=green]{What about the latency of training?}
Our system varies the batch size for a job (within the limits specified by the user) 
in order to maximize the system utilization. 
The range of acceptable batch sizes are typically well known for commonly used models.
In many cases, users determine the range of batch sizes based on the acceptable time for completion of the job.
In such cases, the {\em JSA} can be used to predict the run-time for different batch-sizes.
See section~\ref{app:minmax} for additional discussion on choosing min- and max- batchsizes.

In the autoscaler, new jobs are placed into a temporary holding 
buffer on arrival, in the same order in which they arrive.
The job scalability analyzer processes the jobs from the holding 
buffer. It computes the scaling characteristics 
of the job and then transfers the job back to the autoscaler
buffer from where it can be processed further.

\subsection{Job Scalability Analyzer (JSA)}\label{subsec:ore}~\label{sec:runtimeestimation}

JSA's goal is to 
determine the scaling characteristics of each job, so that the run time of 
the job can be estimated for various $\langle \#~of~GPUs,~batch~size~\rangle$ 
combinations as required.
Certain scaling characteristics are job dependent whereas others are generic,
i.e., dependent on the cluster.
The \JSA\ determines the generic scaling characteristics once at startup 
and repeats it infrequently to account for changes to the cluster hardware.
Job dependent characteristics are determined whenever a new job arrives.

The run-time for a DL training job can broadly be segregated into two components: 
the \emph{processing time} and the \emph{communication time}. 
DL training being data parallel, a batch is typically divided equally amongst 
the GPUs and each GPU works on the part of the batch
allocated to it; we refer to the size of the batch allocated per GPU as ``batch-size-per-GPU''
(computed as batch size divided by number of GPUs).
As the processing happens in parallel over all the GPUs allotted to the job, 
the processing time is 
simply the time required to process batch-size-per-GPU on a single GPU.
The communication time is the time required to reduce the gradients at the end of each iteration;
this is dependent on the number of weights in the model and the number of GPUs.

Thus it suffices for the \JSA\ to estimate: 
(i) for a given job, the processing time for various batch-size-per-GPU values on a single GPU; this is done
by collecting certain job specific scaling characteristics.
(ii) time for communicating different sizes of model weight buffers across a range of GPUs, i.e., the Allreduce-time for various combinations of \# of weights and \# of GPUs in the cluster; this is estimated using generic scaling characteristics.



\subsubsection{Job specific scaling characteristics}~\label{sec:job-specific-scaling-characteristics}
To estimate the processing time:

\begin{enumerate}

\item~\label{step:orefirst} The \JSA\ considers every new job in the temporary holding buffer, executes it on a single GPU for a few iterations with different values of batch-size-per-GPU and records the average time taken per iteration. The methodology for estimating the runtime per iteration is similar to \cite{gandiva}.

\item~\label{step:bspgchoice} The batch-size-per-GPU values are chosen uniformly between $b_{\mathit{min}}$ and $b_{\mathit{max}}$ 
(the maximum batch-size-per-GPU feasible for the job). 

\item These scaling characteristics (average run-time per iteration) are tagged to the 
job metadata and the job is transferred to the temporary autoscaler buffer.

\item \emph{Note}: Later, if the optimizer requires the processing time for 
a value of batch-size-per-GPU outside those considered in Step~\ref{step:bspgchoice}, 
it is determined by interpolating values computed in Steps~\ref{step:orefirst}-\ref{step:bspgchoice}.

\end{enumerate}

Thus, this allows us to estimate the processing time $t^{\mathit{proc}}_j(b_{gpu})$ for a job $j$ when running with
batch-size-per-GPU of $b_{gpu}$.

\subsubsection{Generic scaling characteristics}~\label{sec:generic-scaling-characteristics}
To estimate the AllReduce-time, 
the \JSA\ performs the AllReduce operation over 
\begin{itemize}

\item A range of GPUs $\{ 1,\cdots,k_{\mathit{max}} \}$. 

\item for each number of GPUs  $ k \in \{ 1,\cdots,k_{\mathit{max}} \}$, a range of model weight sizes ($10M, 20M, 30M, \cdots, 100M$ weights).

\end{itemize}

Later, if the communication time is required for other values of
number of weights, it is determined by interpolation of the available values.
We note that it suffices to perform this estimation infrequently 
(it is not required to be done for every job; it can be performed whenever
there are major changes to the cluster). 
Thus, this allows us to estimate the communication 
time $t^{\mathit{comm}}(p,k)$ for a job having $p$ weights
when running on $k$ GPUs.

The data communication time $t^{\mathit{comm}}(p,k)$ for a particular job principally determines how well it scales. 
A higher proportional increase of $t^{\mathit{comm}}(p,k)$ indicates a higher proportion of time is spent in
reducing the gradients and hence less actual processing (or learning) per unit time; such jobs are called \textit{communication bound} and do not scale well (compared to \textit{compute bound} jobs which scale well).

\subsubsection{Estimating the run-time for a job}
As a batch of size $b$ is divided equally amongst GPUs allotted to a DL training job, each GPU processes $\lceil \frac{b}{k} \rceil$
of the input samples. Let

\begin{itemize}
    \item $t^{\mathit{proc}}_j(\lceil \frac{b}{k} \rceil)$ denote the processing time estimate 
for running the job with batch size $\lceil \frac{b}{k} \rceil$ on a single GPU (as described above in Section~\ref{sec:job-specific-scaling-characteristics}).

    \item $t^{\mathit{comm}}(p_j,k)$ denote the communication time estimate for performing the Allreduce operation of $p_j$ weights with $k$ GPUs (as described above in Section~\ref{sec:generic-scaling-characteristics}).
    
\end{itemize}

\noindent The per-iteration run-time for the job is then estimated as:

\[
t^{\mathit{iter}}_j(b,k)=t^{\mathit{proc}}_j(\lceil \frac{b}{k} \rceil)+t^{\mathit{comm}}(p_j,k)
\]

\noindent Hence, the processing rate (number of input samples processed per unit time) 
for job $j$ with batch size $b$ on $k$ GPUs is:

\[
T_j(b,k)=\frac{b}{t^{\mathit{proc}}_j(\lceil \frac{b}{k} \rceil)+t^{\mathit{comm}}(p_j,k)}
\]

In order to make optimal scheduling decisions, the optimizer needs 
the processing rate estimate for a job $j$ when scheduled with a specific batch-size, 
say $b$, and number of GPUs, say $k$. 
This estimate can be computed using the scaling characteristics collected by the \JSA\ 
(this estimation functionality is provided by the \JSA\ for use by the optimizer).

For configurations that are infeasible, 
for instance if batch-size-per-GPU $\lceil b/k \rceil$ does not fit on a GPU,
we take $T_j(b,k)$ to be a large negative number.
This will ensure that the optimizer will never consider this configuration in the optimal solution.

\subsection{Optimizer}~\label{sec:optimizer}
As mentioned before, the goal of the optimizer is to schedule the current set of jobs 
on the cluster GPUs 
in an optimal manner so as to maximize the net throughput (defined as the average number of jobs processed per unit time).

The autoscaler performs a periodic check (at the lapse of every $\triangle$ time interval) whether its temporary input buffer contains some new jobs or whether some existing jobs have completed within the time interval (or both); if any of these conditions are met, the autoscaler adds the currently running job IDs and the new job IDs from the input buffer (in order of arrival) to a list (along with the associated job characteristics of each job) and passes it to the optimizer. 
We next describe in more detail the objective and algorithm of the optimizer.

\subsubsection{Objective}\label{subsubsec:obj}
The optimizer has to determine, for each job, the number of GPUs to be allocated 
to the job and the batch size it should be run with, such that the overall 
throughput of the cluster is maximized. Every job
must be allocated at least $1$ GPU. If it is not possible to do so, 
the optimizer reports that
the problem is infeasible.

The cluster throughput is determined as follows.
For a job, $j$, we treat $T_j(b^{\mathit{max}}_j,1)$ as the baseline processing rate of the job, where
$b^{\mathit{max}}_j$ is the maximum batch-size-per-GPU that can be scheduled for the job.
We now define the {\em throughput scaling factor}, ${\mathcal{T}_j(b , k)}$, 
to be the factor increase in processing rate obtained when the job is run with batch size $b$ and $k$ GPUs
in comparison to the baseline, i.e.,

\begin{equation}\label{sf}
{\mathcal{T}_j(b , k)}=\frac{T_j(b,k)}{T_j(b^{\mathit{max}}_j,1)}
\end{equation}
We shall use the notation $b^{\mathit{opt}}_j(k)$ to denote the optimal batch-size-per-GPU for job $j$ 
when running on $k$ GPUs, i.e., the one that yields the best throughout scaling factor;
this can easily be determined as follows
\begin{equation}\label{eq:bbs}
b^{\mathit{opt}}_j(k) = \arg \underset{b}{\max} \ {\mathcal{T}_j(b , k)}
\end{equation}

Let $J$ be the total number of jobs in the list obtained by the optimizer. 
Then the objective of the optimizer is to determine $k_j$ for $j=1$ to $J$, i.e., the number of GPUs
to be allocated to each job, so as to maximise the total throughput scaling factor of the jobs,
i.e., 
\begin{equation}\label{obj}
\mbox{maximize } \mathcal{T} = \sum_{j=1}^J \mathcal{T}_j(b^{\mathit{opt}}_j(k_j) , k_j)
\end{equation}




\subsubsection{Algorithm} 

Given the objective function (\ref{obj}) for the optimization problem,
a mixed integer program can be formulated to solve the problem. However,
such programs can take very long to solve in practice depending on the number of
jobs and GPUs. We show that the optimal solution to this problem satisfies the
\emph{optimal substructure property} and thus admits a dynamic program. 

{\bf Proof : Optimal Substructure}

The optimal solution to the Optimizer's objective (Equation~\ref{obj}) satisfies the optimal substructure 
property. We shall use the notation $\mathcal{P}(j,K)$ to denote the optimal throughput of the first 
$j$ jobs when allocated a total of $K$ GPUs. 

\begin{proof}
By contradiction. Consider the optimal throughput of the first $j$ jobs when allocated $K$ GPUs, $\mathcal{P}(j,K)$.
Then it is not difficult to see that 
$\mathcal{T}'=\mathcal{P}(j,K)-\mathcal{T}_j(b^{\mathit{opt}}_j(k_j) , k_j)$ must be 
the optimal throughput of the first $j-1$ jobs when scheduled on $K-k_j$ GPUs.
If $\mathcal{T}'$ were not optimal, then a better solution can be 
constructed for the first $j$ jobs on $K$ GPUs by 
combining the allocation of the above solution for the first $j-1$ jobs on $K-k_j$ GPUs
that yields better throughput with an allocation of $k_j$ GPUs to the $j^{th}$ job.
The throughput of this solution would be 
$\mathcal{T}'+\mathcal{T}_j(b^{\mathit{opt}}_j(k_j) , k_j) > \mathcal{P}(j,K)$
which would contradict the optimality of $\mathcal{P}(j,K)$.
\end{proof}




Thus we can formulate a dynamic program (DP) to compute the optimal solution 
\textit{every time} the autoscaler consults the optimizer by using  
a DP table for $\mathcal{P}(\cdot,\cdot)$ and populating the entries of
this table iteratively using the following relation:
\begin{equation}\label{eq:dp}
\mathcal{P}(j,K) = \underset{1 \leq k \leq k_{\mathit{max}}} {{\max}} \ \
[ \mathcal{P}(j-1,K-k) + \mathcal{T}_{j}(b^{\mathit{opt}}_{j}(k) , k) ]
\end{equation}

The optimal allocation $k_j$ to the $j^{th}$ job is thus given by: 
\begin{equation}\label{eq:opt_gpus}
k_j = \underset{ 1 \leq k \leq k_{\mathit{max}}}  {\arg{\max}} \ \ 
[ \mathcal{P}(j-1,K-k) + \mathcal{T}_{j}(b^{\mathit{opt}}_{j}(k) , k) ]
\end{equation}

A practical way to solve the \textit{dynamic program} (DP) in (\ref{eq:dp}) is to initialize the array $\mathcal{P}(\cdot,\cdot)$ to a large negative number and build the DP table progressively from $j =1, k = 1$ to $j =J, k = K$. Note from (\ref{eq:dp}) that the optimizer might return infeasible for such a problem setting, which will mean that after the completion of the DP, $\mathcal{P}(J,K) \leq 0 $. In such a case, the autoscaler (next section) is responsible for figuring out a feasible solution by appropriately trimming the input job-queue to the optimizer.

\begin{algorithm}
\caption{OPTIMIZER}
\label{algo:dp}
\begin{algorithmic}
\small
\REQUIRE exec-queue $\mathcal{E}_t$, total GPUs $K$
\RETURN Updated $\mathcal{E}_t$ with new GPU allocation, status
\STATE $ J \leftarrow \text{LENGTH}(\mathcal{E}_t)$
\STATE $\mathcal{P} \leftarrow \mathbbm{-\infty}_{J \times K}$ \COMMENT{initialize $J\times K$ array of $-\infty$}
\STATE $SOL \leftarrow 0_{J \times K}$ \COMMENT{initialize $J\times K$ array of $0$s}
\STATE $\mathcal{P}(0,:) \leftarrow 0$ \COMMENT{no jobs mean zero utilization}
\FOR{$j = 1,\cdots,J$}
\FOR{$k = 1,\cdots,K$}
\FOR{$g = 1,\cdots,k_{max}$}
\IF{$k-g \geq 0$}
\STATE $p \leftarrow \textsc{JSA.recall}(\mathcal{E}_t(j),k) ~+~ \mathcal{P}(j-1,k-g+1)$
\IF{$p > \mathcal{P}(j,k)$} 
\STATE $\mathcal{P}(j,k) = p~~;~~SOL(j,k)=k$ \COMMENT{better utilization found}
\ENDIF
\ENDIF
\ENDFOR
\ENDFOR
\ENDFOR
\IF{$\mathcal{P}(J,K) > 0$}
\STATE $status \leftarrow ``feasible" ~~~;~~~ j \leftarrow J ~~~;~~~ k \leftarrow K $
\WHILE{$j > 0$}
\STATE $\textsc{updateGPU}(\mathcal{E}_t(j),SOL(j,k))$ \COMMENT{set GPU allocation for job $j$}
\STATE $j \leftarrow j - 1~~;~~ k \leftarrow k - SOL(j,k)$
\ENDWHILE
\ELSE
\STATE $status \leftarrow ``infeasible"$
\ENDIF
\end{algorithmic}
\end{algorithm}

\fxfatal{what do we do in this case? let the allocations unchanged. Also need to link with job dropping mention by autoscaler in the next subsection}

The implementation of the dynamic program in (\ref{eq:dp}) is shown in algorithm (\ref{algo:dp}). 
The entries $\mathcal{P}(0,K)$ of the $DP$ table are initialized to $0$ as the throughput is $0$ for $0$ jobs
irrespective of the number of GPUs. All other entries of the $DP$ table 
are initialized with a large negative number.

The solution array $SOL(j,k)$ denotes the optimal GPUs allocated to the $j$-th job when the total GPUs allocated to all the $j$ jobs is $k$. 

$\mathcal{E}_t$ is the input job queue that is sent to the optimizer. The function call \textsc{JSA.recall()} looks up the value of the throughput scaling factor $\mathcal{T}_j(b^{\mathit{opt}}_j(k) , k)$ for the $j$-th job $\mathcal{E}_t(j)$ in the job-queue when run on $k$ GPUs. 

It is easy to see that the run-time complexity of the dynamic program is $O(JKk_{max})$
based on the three for loops. A problem will only be feasible if the number of jobs is
no more than the number of GPUs. Thus, we can assume that $J \le K$. 
Therefore, even for $400$ GPUs and $k_{max}=10$, the complexity is no more than an order of $2$M operations (\emph{milliseconds} on modern CPUs).

If a feasible solution exists ($\mathcal{P}(J,K) > 0$), each job in the queue will have their respective GPU allocation field updated to the optimal number of GPUs with a call to \textsc{updateGPU()}. In case no feasible solution exists, an 'infeasible' status is sent back to the autoscaler and the existing GPU allocation field for all the jobs in the input queue is left untouched.


\subsection{Autoscaler}~\label{sec:autoscaler}

\begin{figure}[htb]
\hrule
\begin{distribalgo}[1]
\small
\vspace{1mm}
\INDENT{\textbf{init}}
\STATE{$\executing[]$}\COMMENT{Jobs currently executing in DL platform}
\STATE{$\arrived[]$}\COMMENT{Jobs that arrived since last scaling action}
\STATE{$\finished[]$}\COMMENT{Jobs that completed/failed since last scaling action}
\STATE{$\trial[]$}

\ENDINDENT

\myalgospace

\UPON {$\arrival{\job[]}$}
\STATE{$\arrived \leftarrow \enqueue{\arrived}{\job[]}$}
\STATE{$T_j[][] \leftarrow \JSA.\process{\job[]}$}\COMMENT{Get various values of $T_j(b,k)$ from the \JSA}
\STATE{$\textsc{AddToMetadata}(\job[],T_j[][])$} 
\ENDUPON

\myalgospace

\UPON {$\departure{\job[]}$}
\STATE{$\finished \leftarrow \finished \cup \set{\job}$}
\ENDUPON

\myalgospace

\UPON{$\textsc{MakeScalingDecisions}()$} 
\IF{($\arrived \neq \set{}~||~\finished \neq \set{}$)}
\STATE{$\executing \leftarrow \executing \setminus \finished$}
\STATE{$i \leftarrow 1$}
\WHILE{$i \leq \textsc{Length}(\arrived)$}
\STATE{$\trial \leftarrow \executing \cup \arrived[i]$} \COMMENT{Add one by one} 
\STATE{$allocation,possible \leftarrow \textsc{Optimize}(\trial)$}
\IF{$possible$}
\STATE{$\executing \leftarrow \trial$}
\STATE{$\arrived \leftarrow \remove{\arrived}{i}$}
\ENDIF
\STATE{$ i \leftarrow i + 1$} \COMMENT{Go to the next job in \arrived}
\ENDWHILE
\STATE{$\textsc{SendToDLPlatform}(allocation, \executing)$}
\ENDIF
\ENDUPON
\end{distribalgo}
\hrule

\caption{High-level Pseudocode of Autoscaler}~\label{algo:autoscaler}
\end{figure}

The autoscaler (pseudocode in Figure~\ref{algo:autoscaler}) maintains a list ($\executing$) of jobs that are
currently being executed by the DL platform. 
It queues all arriving jobs in another list ($\arrived$). Each job in the 
queue is first processed by the \JSA. The \JSA\ computes
the scaling characteristics of the job as described in 
Section~\ref{sec:runtimeestimation} and adds these
characteristics to the job metadata.  The autoscaler is 
also notified by the DL platform whenever a job completes
(or fails due to user error) and stores these jobs in a list ($\finished$),
removing them from the $\executing$ list before invoking the optimizer. To be clear,
the optimizer will be invoked even if no new job arrives, but 
jobs leave.

Ideally, the autoscaler should invoke the optimizer
every time a job arrives or leaves. However, in practice,
this can lead to \emph{thrashing} -- a significant amount 
of time is spent scaling DL jobs up and down through
checkpoint/resume that it affects the progress made by the DL job.
Consequently, in practice, the autoscaler consults the optimizer
periodically (say every $\triangle$ minutes). $\triangle$ is typically based
on the amount of time a DL job can be queued without affecting 
user experience and expectations. See section~\ref{app:delta} for additional discussion on choosing $\triangle$ value.


Every $\triangle$ minutes, the autoscaler invokes the optimizer to determine
(i) whether GPUs can be allotted to as many newly arrived jobs (from $\arrived$) and
(ii) how many GPUs are to be allotted to each job. This is done in an 
iterative and incremental manner -- first by removing jobs that have terminated,
and then trying to add jobs from $\arrived$ one by one until the 
optimizer returns 'infeasible'. Once the set of jobs that can be executed
and the allocations to said jobs are determined, the autoscaler 
interacts with the DL platform to spawn new jobs, and scale existing 
jobs up or down. The autoscaler can also be \emph{modified to drop (i.e., reject) pending jobs} 
(which are not feasible to accommodate) after Step 23 if
queueing is undesirable. 

\subsection{Simulator Design and Implementation}~\label{sec:simulator}
We develop a simulator in order to evaluate our techniques on large cluster settings;
given the difficulty in getting access to large GPU-enabled clusters for large periods of time.
The simulator is based on the discrete event simulation (DES) \cite{des} methodology, 
wherein the simulation is driven by time based events.
An event is created in the simulator whenever a job arrives or completes. 

The input to the simulator is a job arrival file that contains meta-information for the jobs
along with their arrival time. The meta-information includes the job details 
(such as maximum batch size, minimum batch sizes, number of epochs to process, etc.)
along with the job specific scaling characteristics collected by the job scalability analyzer (JSA)
(see Section~\ref{subsec:ore}).
Recall that the {\em JSA} runs the jobs 
for only a few iterations on up to $k_{max}$ GPUs to collect the job specific scaling data.

The simulator reads the job arrival information from the file to create arrival 
events for the jobs based on their arrival time. 
Based on the schedule returned by the optimizer, the simulator updates the GPU allocations to the jobs
and updates their completion time based on the run-time estimates as determined in Section~\ref{subsec:ore}. 
It also adds a completion event for any new job that has been scheduled.
In case the optimizer returns that no allocation is feasible, then the newly arrived job is 
either dropped or put into a queue depending on the configuration of the simulator; the simulator
can be run in both modes -- with and without queuing. 
In case of queuing, the first job from the queue is considered for execution on the next
job completion event; it is treated like a fresh job and the schedule is determined accordingly.
The simulator then proceeds to process the next event until all the events are processed.





\section{Experimental Evaluation}~\label{sec:eval}

\begin{table*}[htb]
\centering
\scalebox{1.0}{
\begin{tabular}{ |c|c|c|c|c|c| } 
 \hline
{\bf  Category} & {\bf Dataset} & {\bf Model} & {\bf Weight Size} & {\bf Min BS, Max BS} & {\bf Characteristics} \\  \hline
 1 & CIFAR100 & resnet50 & 24M & 32,256 & Elastic, Compute Bound \\ \hline
 2 & CIFAR100 & alexnet & 58M &  16,256 & Elastic, Communication Bound \\ \hline
 3 & CIFAR100 & vgg11\_bn & 10M & 16,1024 &  Elastic, Balanced \\  \hline
 4 & Food101 & alexnet & 58M & 128,128 & No Elasticity \\
 \hline
\end{tabular} } 
\caption{Job Category Information\fxfatal{why 2 is elastic and 4 is not is not clear from this table}}
\label{eval:table_category}
\end{table*}

We evaluated our elastic scaling technique using a real 40 GPU cluster.
Each of the machines was equipped with dual Intel Xeon E5-2620 v4 (2.10 Ghz) with a total of 16 CPU cores,
64 GB RAM, 10 GbE public and private uplinks and 2 P100 GPUs. The cluster was managed by Kubernetes v1.12, and 
had \FfDL~\cite{ffdl} installed as the DL platform used to execute jobs. We used PyTorch~\cite{pytorch} for training the jobs.

Additionally, we also evaluated our elastic scaling techniques using the simulator described in Section~\ref{sec:simulator} over a 400 GPU cluster. Throughout the rest of this section, experiments conducted on the actual cluster
 will be tagged with the ``Cloud'' label when results are
presented or plotted, while experiments conducted through the simulator will be tagged with the ``Simulator'' label.

\begin{table}[htb]
\centering
\begin{tabular}{ |c|c|c|c|c|c| } 
 \hline
{\bf Batch size per GPU} & 8 & 11 & 16 & 22 & 32 \\ \hline
{\bf Scaling Factor} & 0.86 & 1.06 & 1.3 & 1.45 & 1.66 \\ \hline
\end{tabular} 
\caption{Throughput Scaling factors for category 1  jobs on 2 GPUs}
\label{table:sf_cat1_2GPU}
\end{table}


\begin{figure*}[htbp]
\centering 	
	
	\subfigure[Category 1: Random-BS baseline,  High Arrival]
	{
		\includegraphics[width=0.4\textwidth, keepaspectratio=true]{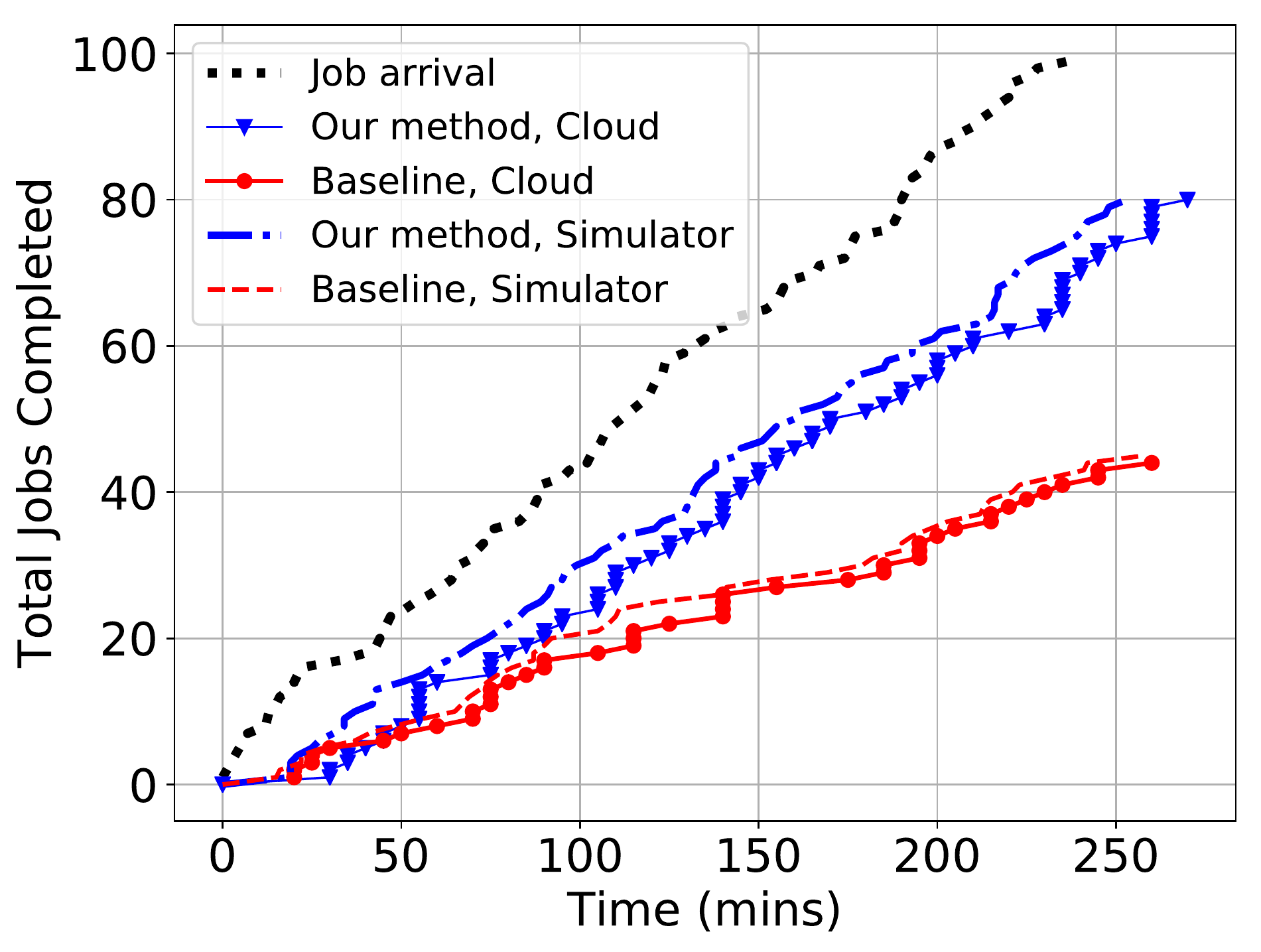}
		\label{plot_mb_job_completed_randomBS_cat3}
	} 
	~
	\subfigure[Category 2: Random-BS baseline,  High Arrival]
	{
		\includegraphics[width=0.4\textwidth, keepaspectratio=true]{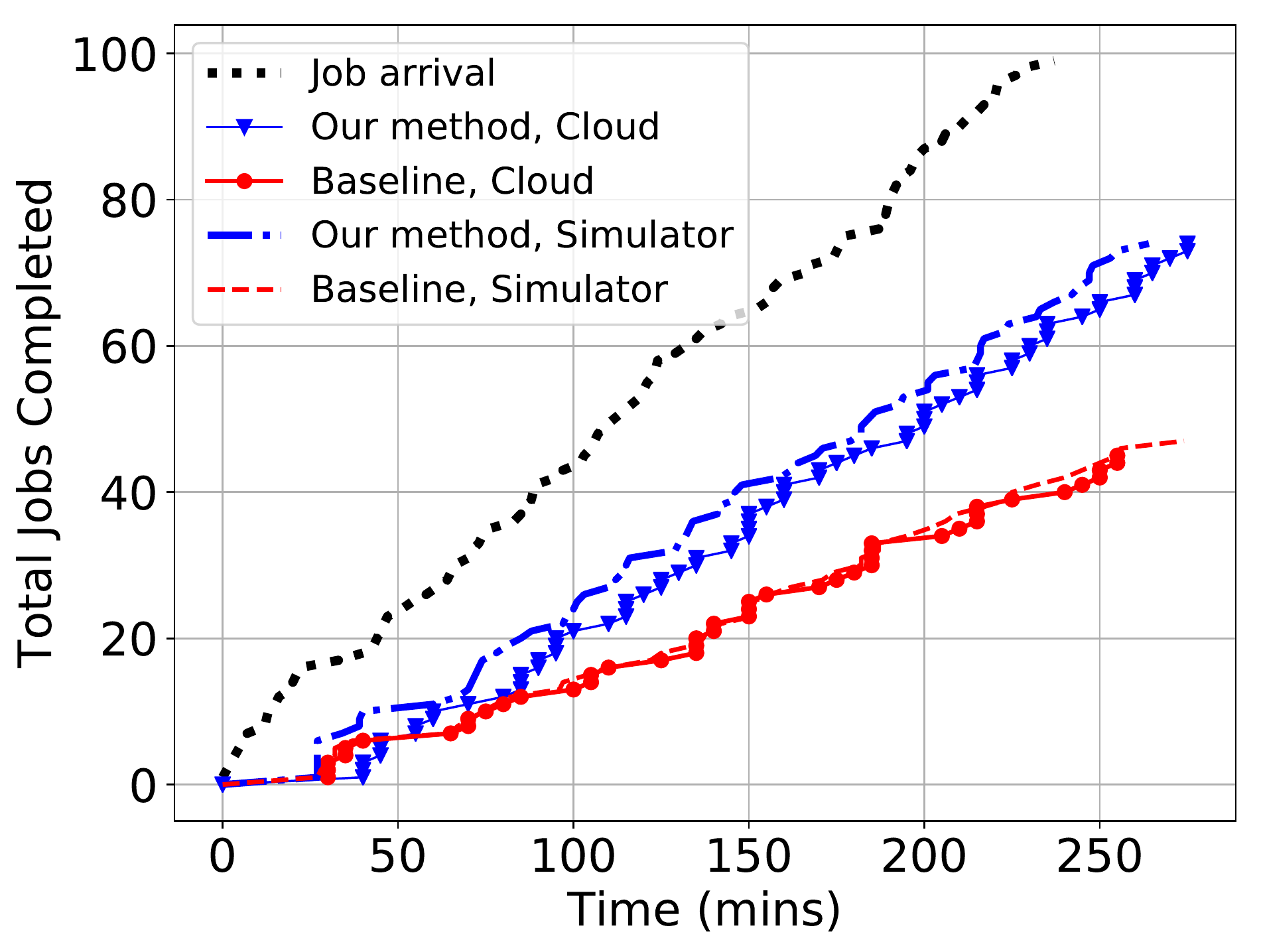}
		\label{plot_mb_job_completed_randomBS_cat2}
	}
	~
	\subfigure[Category 3: Random-BS baseline,  High Arrival]
	{
		\includegraphics[width=0.4\textwidth, keepaspectratio=true]{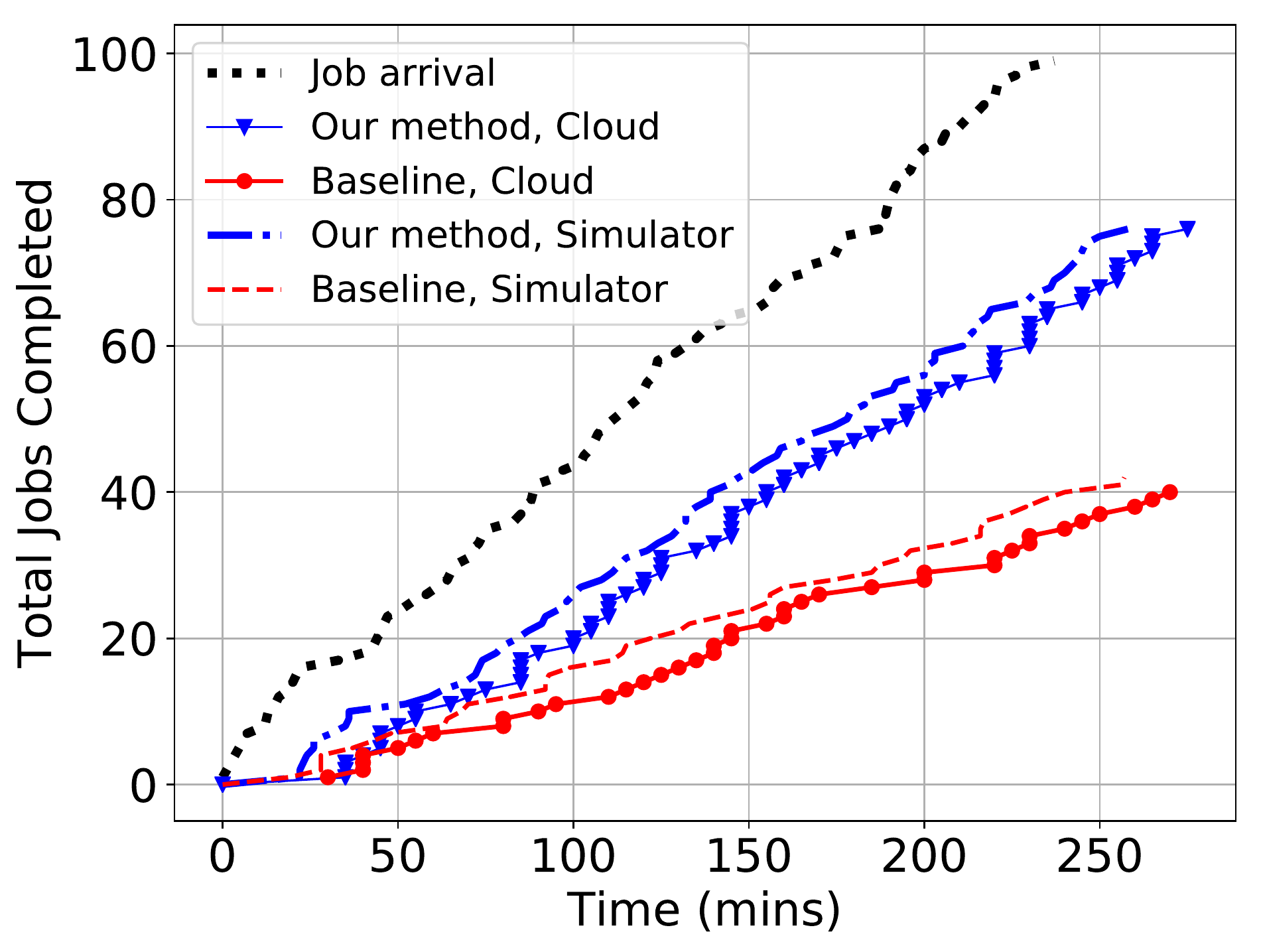}
		\label{plot_mb_job_completed_randomBS_cat1}
	}
	~
	\subfigure[Category 4: Fixed Batch Size,  High Arrival]
	{
		\includegraphics[width=0.4\textwidth, keepaspectratio=true]{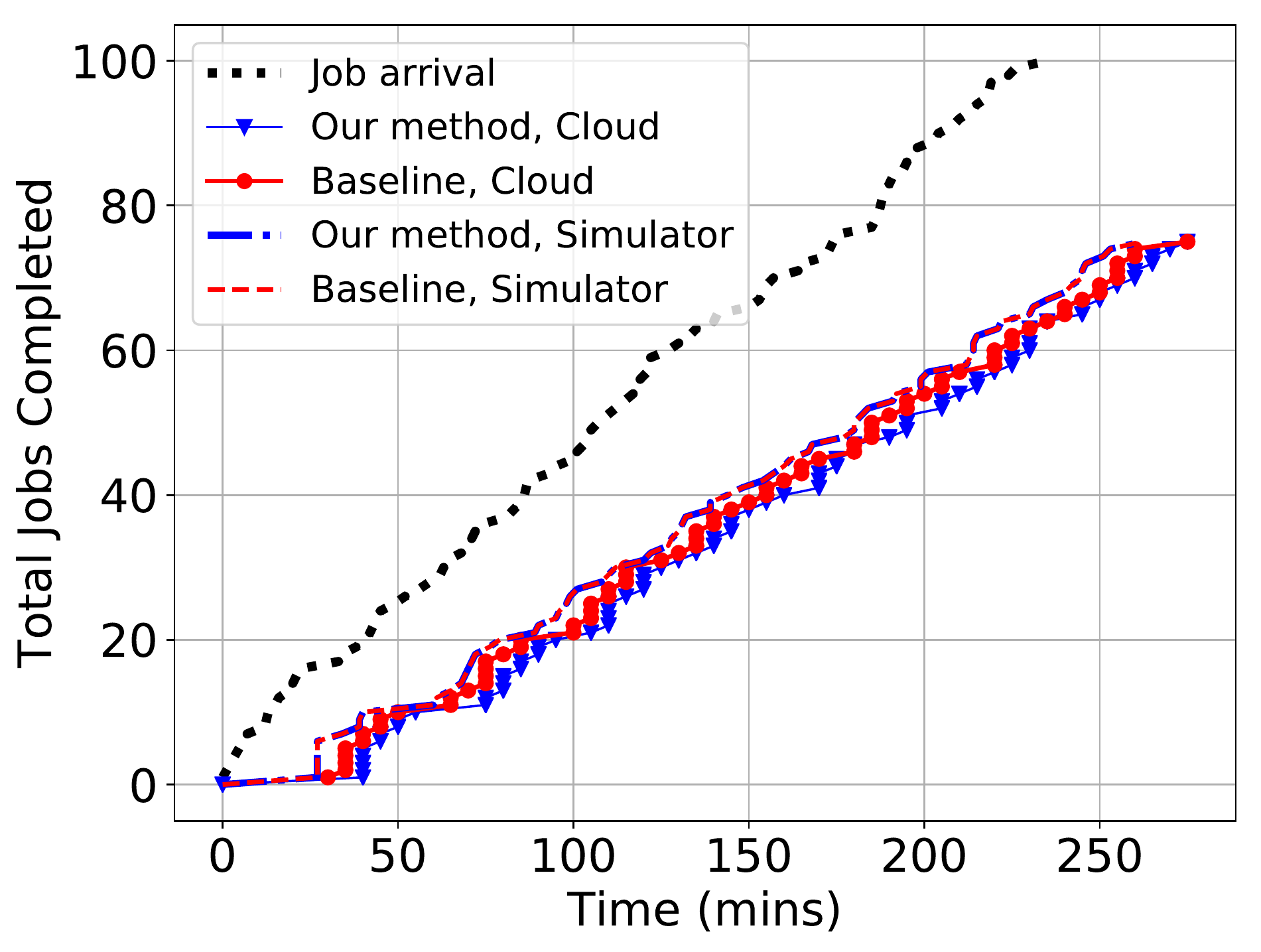}
		\label{plot_mb_job_completed_fixedBS_cat4}
	}

\caption{
Effect of job categories.}%
\label{res:fig_mb_2}%
\label{fig:2}
\end{figure*}

\begin{figure}[htb]
\centering 
\subfigure[Category 1: Random-BS Baseline - Low Job Arrival]
{
	\includegraphics[width=0.4\textwidth, keepaspectratio=true]{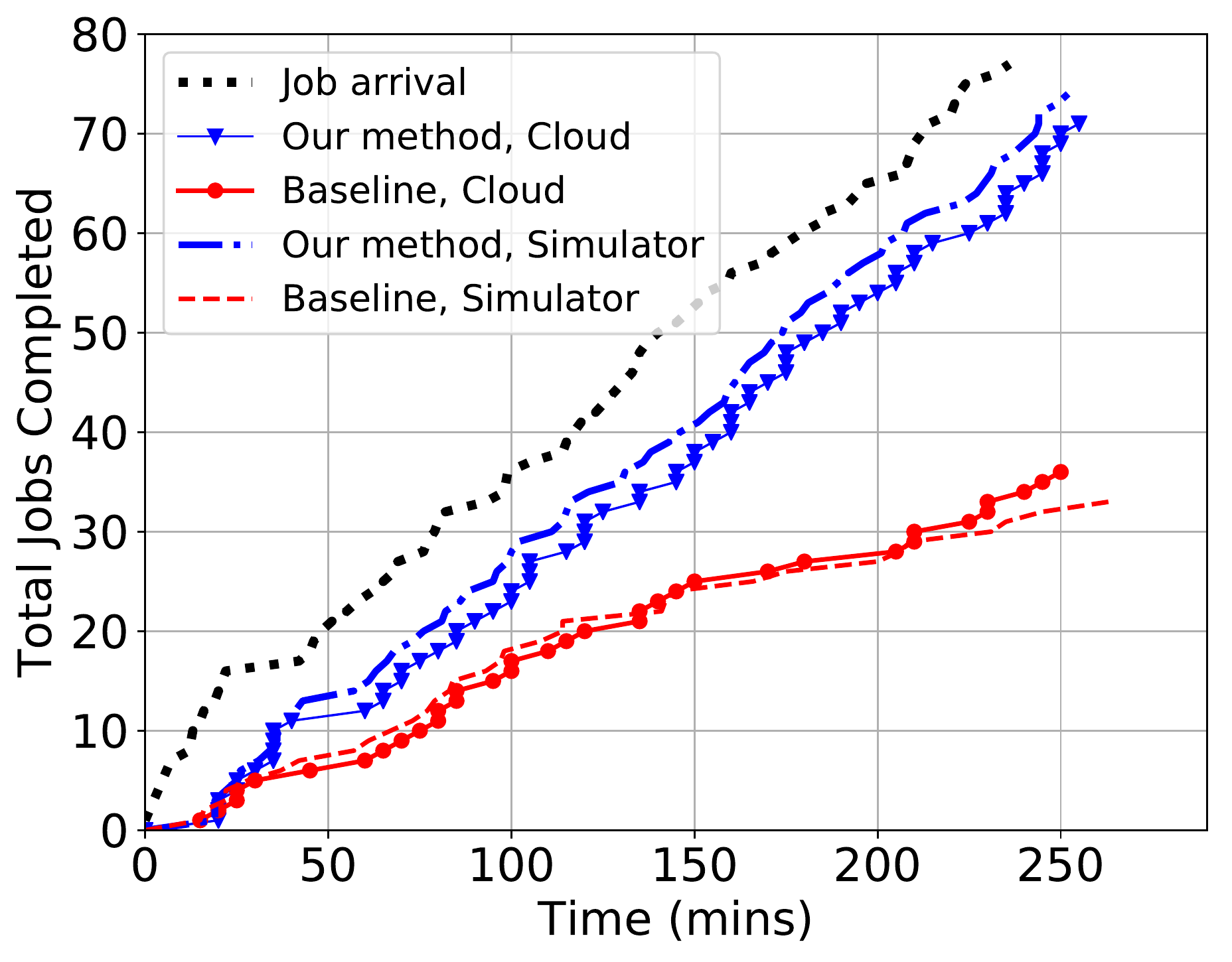}
	\label{plot_mb_job_completed_randomBS_light}
}
~
\subfigure[Category 1: Random-BS Baseline - Bursty Job Arrival]
{
	\includegraphics[width=0.4\textwidth, keepaspectratio=true]{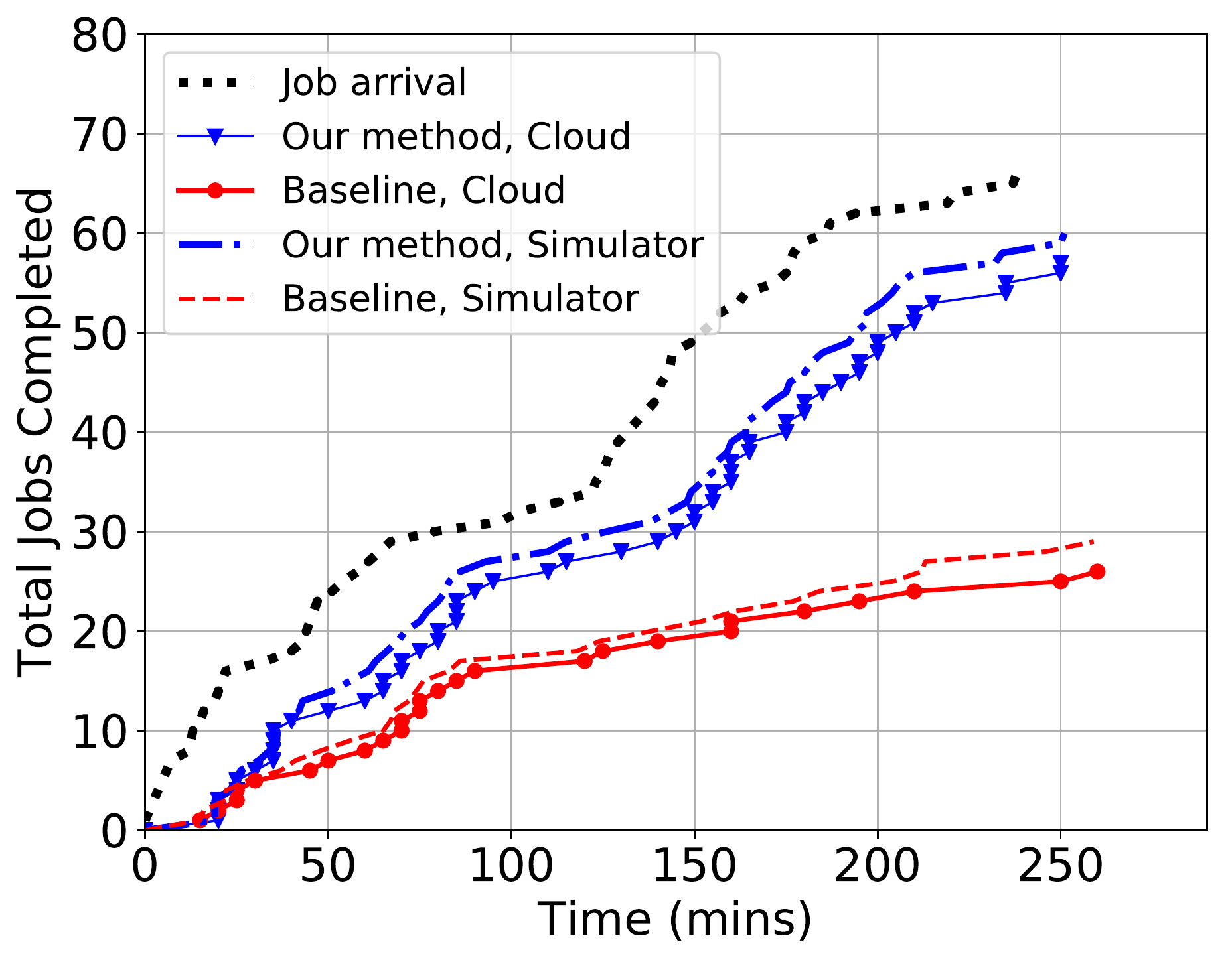}
	\label{plot_mb_job_completed_randomBS_bursty}
}
\caption{
Effect of Arrival Patterns. 
}%
\label{res:fig_mb_3}%
\label{fig:3}
\vspace{-5mm}
\end{figure}

\begin{figure}[htb]
\centering
	\subfigure[t][Allow Dropping of Jobs (No Queue)]
	{
		\includegraphics[width=0.4\textwidth, keepaspectratio=true]{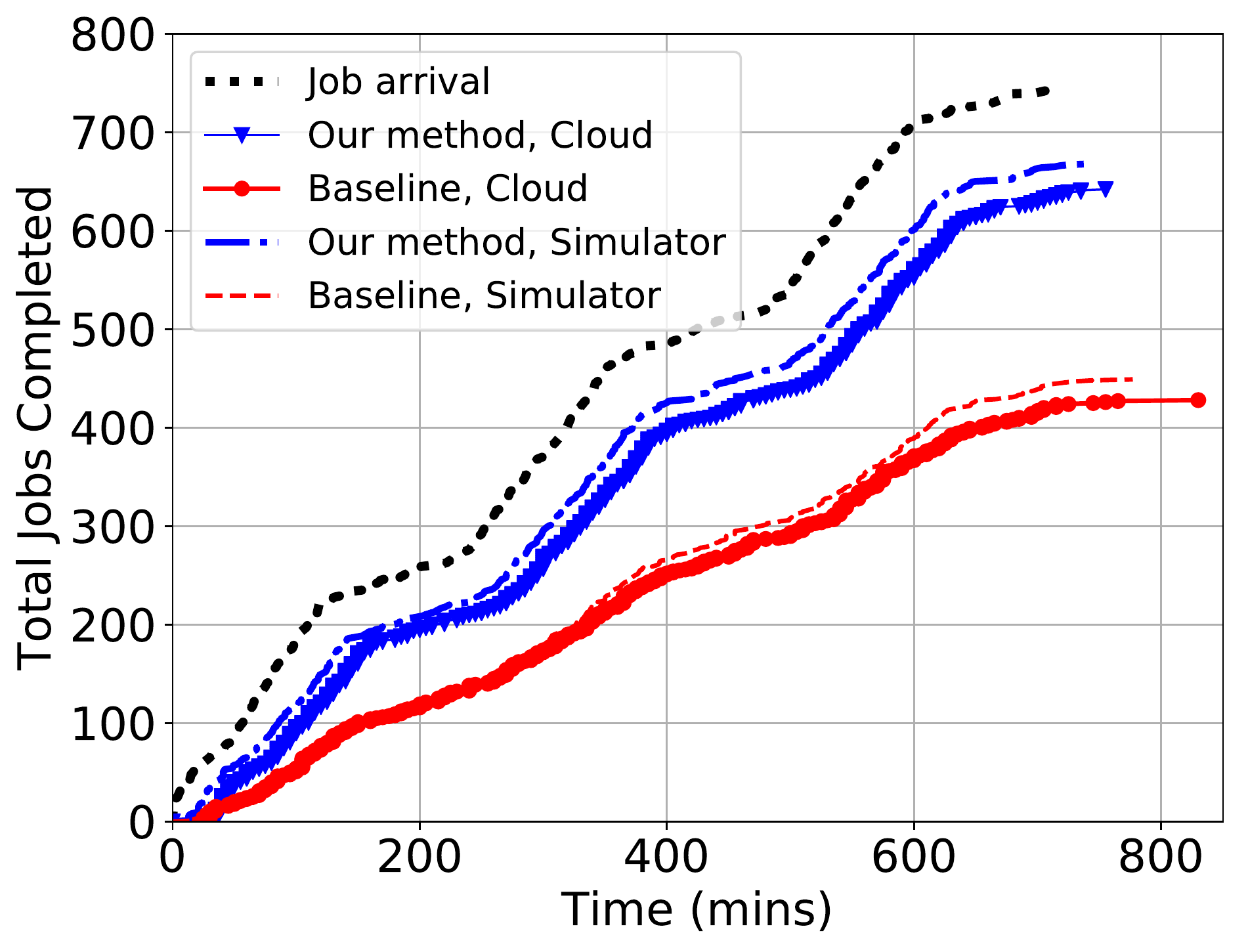}
		\label{plot_40GPU_job_completed_bursty}
	}
    \\
	\subfigure[t][No Dropping of Jobs (Queuing of Jobs)]%
	{
		\includegraphics[width=0.4\textwidth, keepaspectratio=true]{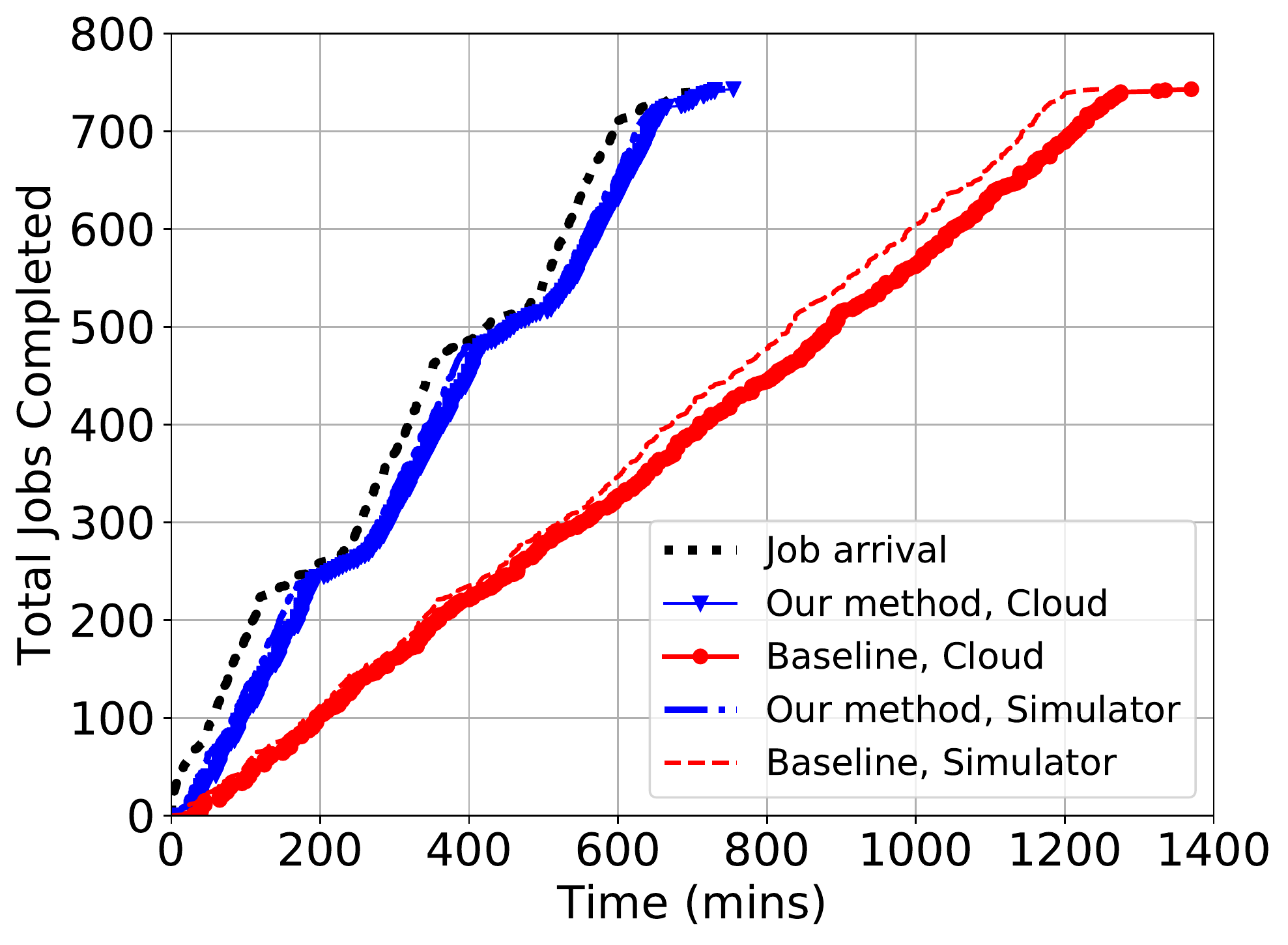}
		\label{plot_40GPU_job_completed_bursty_queue}
	}%
	
\caption{
Effect of Queueing (Random-BS baseline)\fxfatal{need to use character plots to differentiate in grey scale}
}%
\label{res:fig_40GPU_bursty}%
\label{fig:40}
\vspace{-5mm}
\end{figure}




\subsection{Benchmarks : Job Categories and Arrival}~\label{sec:jobcategories}~\label{eval:microbenchmarks}~\label{sec:benchmarks}


The neural network model and the dataset of a job determines the communication vs compute characteristics of the job. 
A compute bound job scales well with larger number of GPUs compared to a communication bound job which scales poorly 
due to communication overhead - this has been captured in the throughput scaling factor calculation for a job in Section \ref{subsubsec:obj}. 


Since the performance of our elastic scaling techniques will principally depend on the distribution of the throughput 
scaling factors of the jobs arriving in the arrival queue, we focus on four categories - (1) jobs with lower throughput scaling factors (communication bound), (2) jobs with 
higher throughput scaling factors (compute bound), (3) jobs with balanced throughput scaling factors 
(compute-communication balanced) and (4) jobs with no elasticity (not intended to scale, will run on fixed number of GPUs).

We have chosen four representative jobs belonging to the four categories, the details of which are given in Table \ref{eval:table_category}. Sampling jobs from these four categories randomly, we reliably generate a real world DL job arrival scenario that has a realistic distribution of jobs with different scaling potentials. Each benchmark only runs jobs of a single job category;
all jobs have the same length (execution time)
irrespective of the category; it is set such that each job takes about half an hour
to execute on a single GPU with the maximum feasible batch size. 
The jobs in a benchmark arrive in one of three patterns: low arrival, high arrival
or bursty arrival over a 240/480/720 minute period (will be clear from context).

The job arrivals have been generated by sampling from a Poisson distribution 
with mean job arrival rate as the design parameter. We define the job arrival rates as follows. 
A base job arrival rate $\lambda$ is defined as the expected completion rate (i.e., reciprocal of 
expected completion time) of a job sampled uniformly from the 4 categories, on a single GPU with 
maximum batch size per GPU. A \textit{high} job arrival rate signifies that the mean arrival 
rate of the Poisson distribution has been set to $k_{max} \lambda$ where $k_{max}$ is the 
maximum GPUs allocated to each job. A \textit{low} arrival rate sets the mean of the 
Poisson distribution to $\frac{\lambda k_{max}}{4}$. A \textit{bursty} arrival rate alternates 
the mean of the Poisson distribution between high and low (defined previously) every 60 or 120 mins as required by the experiment.

Furthermore, the jobs in a benchmark baseline run may be scheduled with batch-sizes following
one of three settings: maximum batch-size (Max-BS) for that category, minimum batch-size (Min-BS) for that category, and random batch-size (Random-BS) which is a value picked uniformly randomly between the Max-BS and Min-BS for that category.
A benchmark is run for a chosen combination of job category (1, 2, 3 or 4), job arrival pattern (high, low, bursty) and batch-size setting in case of baseline (Mas-BS, Min-BS, Random-BS).


\subsection{Choosing the Baseline}


A completely non-elastic scheduler can force a constant total batch size on a fixed number of GPUs through the job lifetime; however, that is evidently a weak baseline. 

For a stronger baseline, we fix the total batch size of a job to be a value selected from the range of allowed batch sizes for that job category, but we allow the total batch of the job to be distributed over several GPUs; this models a traditional elastic scaler and does not take into account the characteristic (batch size range) of the DL job. The allocation of GPUs to baseline jobs is done through the same optimizer that is used in our elastic scaling technique. The optimizer is invoked periodically (every $\triangle$ minutes) to allocate the relevant number of GPUs for baseline jobs based on their throughput scaling factors such that the total batch size remains fixed across the allocated GPUs. \emph{To reiterate, our baseline is also elastic in terms of total number of GPUs but doesn't vary the total batch size.} 



\subsection{Metrics}~\label{eval:metric_def}
\todo[inline, color=green, size=\tiny]{Why don't we use Samuel Kounev's metrics? Why not SPEC metrics?}

Unfortunately, existing metrics for evaluating elasticity~\cite{spec-iaas-2018, elasticrmi, Ilyushkin-TOMPECS-autoscaler-eval} have been designed for and apply only to cluster elasticity.
We therefore define new metrics based on:

\begin{itemize}
    \item {\bf  Optimal GPU time for all scheduled jobs \newline (Opt\_Sch\_Time)}. This is defined as the sum of job lengths (time) for all the scheduled jobs on a single GPU. This is the minimum GPU-time needed to complete all the scheduled jobs (GPU-time means actual runtime on GPU, and we expect imperfect scaling).
    
    \item {\bf Actual GPU time for all the scheduled jobs \newline (Act\_Sch\_Time)}. This is defined as the sum of [ (actual number of GPUs used for running the jobs) x (Time duration for which these GPUs were used) ]

\end{itemize}

For example, if a single scheduled job requires 10 mins to complete on a single GPU but requires 6 mins to complete on 2 GPUs, then its Optimal GPU time will be 10 mins whereas the Actual GPU time will be 12 mins (=6x2).

We measure and report the performance in terms of two metrics (i) {\bf Scheduled Job Scaling (SJS) Efficiency}, defined as \textit{Opt\_Sch\_Time / Act\_Sch\_Time}, and (ii) {\bf Job Drop Ratio}, defined as \textit{Number of jobs dropped / Total number of jobs}. The first metric shows the average scaling efficiency of all the scheduled jobs i.e. how well the scheduled jobs scale across GPUs. The second metric shows the proportion of jobs that have been dropped.



\subsection{Comparison with Max-BS and Min-BS}
\label{result:compare_min_max}
We first compared our elastic scaling algorithm with two simple strategies:
one which schedules with the maximum batch-size, {\em Max-BS}, specified by the user,
while the other schedules with the minimum batch-size, {\em Min-BS}.
Note that both these strategies can use the baseline algorithm for scheduling as
they do not dynamically change the batch-size of the jobs.

\begin{figure*}[htb]
\centering
	\subfigure[Jobs Completed using Baseline with Max Batch Size - High Job Arrival. Job Categories 1 and 2]
	{
		\includegraphics[scale=0.45]{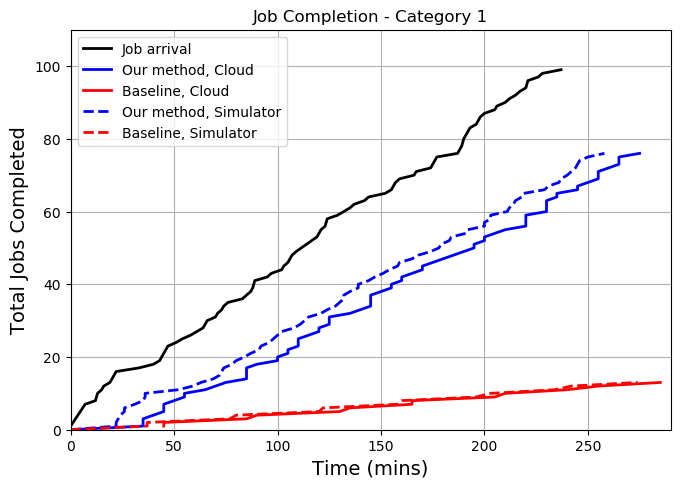}
		\includegraphics[scale=0.45]{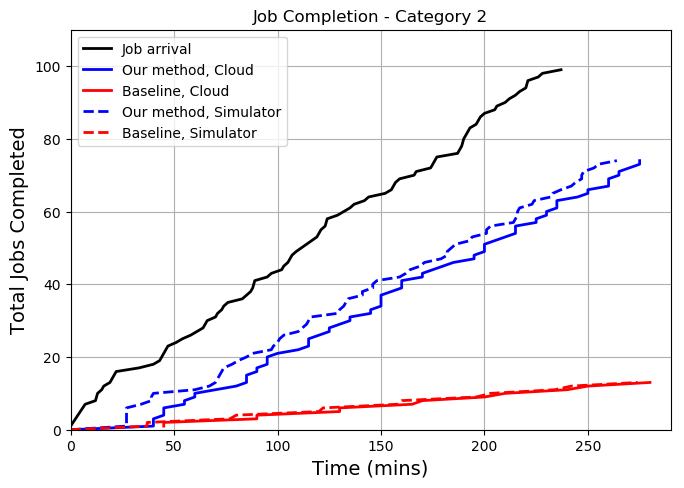}
		\label{plot_mb_job_completed}
	}
	\subfigure[Jobs Completed using Baseline with Max Batch Size - High Job Arrival. Job Categories 3 and 4]
	{
		\includegraphics[scale=0.45]{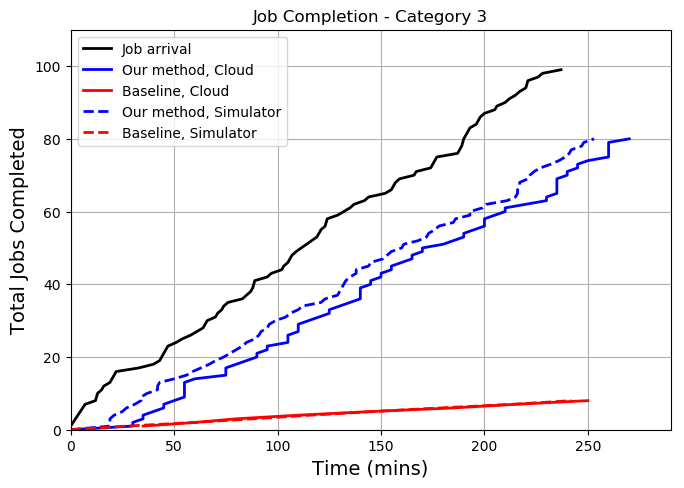}
		\includegraphics[scale=0.45]{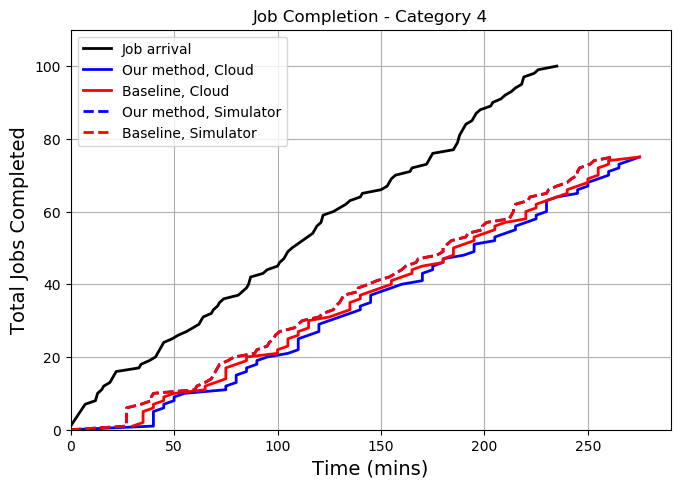}
	}

	\subfigure[Jobs Completed using Baseline with Min Batch Size - High Job Arrival]
	{
		\includegraphics[scale=0.33]{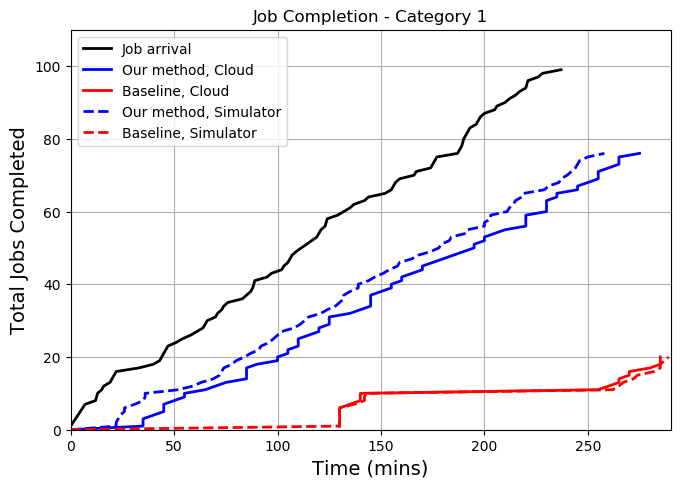}
		\includegraphics[scale=0.33]{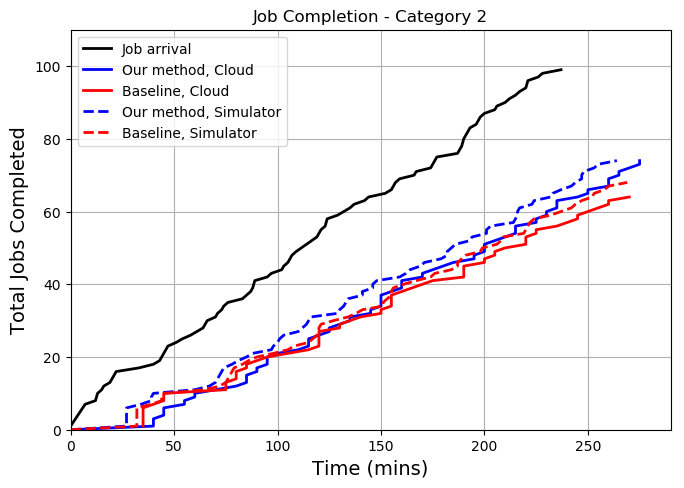}
		\includegraphics[scale=0.33]{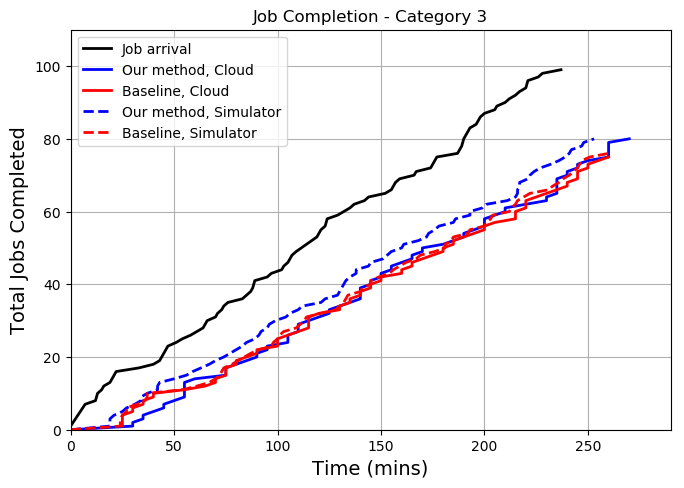}
		\label{plot_mb_job_completed_minBS}
	}
\caption{
Comparison with Max- and Min-BS. Job arrival described in Section~\ref{eval:microbenchmarks}.}%
\label{res:fig_mb_1}%
\label{fig:1}
\end{figure*}

Figures~\ref{fig:1}(a-c) compare our elastic scaling technique with the baseline,
by plotting the number of jobs completed over time, while running Category 1 jobs with high and low arrival rate
(cf. Section~\ref{sec:jobcategories} \& \ref{eval:microbenchmarks}). The first thing we
observe is how closely the simulator results match the execution on the real 40 GPU cluster (``Cloud'').
Figure~\ref{fig:1}(a) illustrates the performance benefits vis-a-vis the Max-BS baseline and high arrival rate.
It can be seen that the elastic scaling algorithm performs significantly better than the 
baseline algorithm, scheduling $\approx$ 10 $\times$ more jobs in comparison. This is because, with high arrival rate,
the baseline is able to schedule only a limited number of jobs as the batch-size is high and thus
the jobs cannot be scaled down to a small number of GPUs. The elastic scaling algorithm on the other
hand can reduce the batch size of the job itself to fit on a small number of GPUs; if the
minimum batch-size allows, it can even schedule the jobs on $1$ GPU. 
Thus, it minimizes the number of
dropped jobs.

Figures~\ref{fig:1}(b-c) illustrate the performance differential while using the Max-BS baseline
for high and low arrival rates respectively. From Figure~\ref{fig:1}(b), we observe that the performance of the
baseline algorithm is quite comparable to that of the elastic scaling algorithm
as both work with small batch sizes with high arrival rate. 
However, the baseline algorithm
does not perform as well with low arrival rate in Fig~\ref{fig:1}(c). 
We observe that the elastic scaling algorithm
does marginally better than the baseline algorithm in terms of jobs completed.
The average job completion time for the elastic scaling algorithm is 
23.04 mins whereas for the baseline algorithm, it is 27.39 mins. 
Thus jobs complete 16\% faster using
the elastic scaling algorithm. This is explained by the fact that with low arrival rate,
both the algorithms try to maximize the GPUs used; however, the elastic scaling algorithm
is able to increase the batch size as it scales to larger number of GPUs whereas the
baseline algorithm cannot. As can be expected, the scaling behaviour is better when there
is more work to be done by every GPU, and thus our algorithm is able to achieve 
better scaling resulting in quicker completion of the jobs.

To further support this argument, we conducted another experiment where we just ran a single 
category 1 job on a 5 P100 GPU setup. 
In one experiment, we ran with the elastic scaling algorithm and in another
we ran with the baseline algorithm using the minimum batch-size.
We observed that with the elastic algorithm,  the job completed 
1.6X faster than with the baseline algorithm;
the baseline algorithm used a batchsize of 32 on 2 GPUs (16 per GPU) whereas the elastic scaling 
algorithm used a batch size of 160 on 5 GPUs (32 per GPU).

\todo[inline, color=green]{Not addressing : Using a fixed batch size can lead to faster convergence, i.e., convergence in fewer epochs. By dynamically varying batch size, aren't you running the job for more epochs than needed?}

\todo[inline]{Why invoke the optimizer every 15-30 minutes in a real run?}

\todo[inline, color=green]{Not Addressing: CNNs have high compute-to-communication ratio and hence more amenable to scaling. A fully-connected architecture has the opposite characteristic, hence will show comparatively lower but significant benefits. We're happy to add a discussion on this.}

\todo[inline, color=green]{Not Addressing: The non-sparse updates (as in our autoscaler) represent the worst case scenario (scaling-wise) for our algorithm, and thus sparse updates can only make our approach more efficient. We are working on extending the JSA to incorporate sparsity-aware estimation.}

In practice, however, most users do not run with Max- or Min- batch size
due to cost and time considerations. 
One notable observation through these results is that the simulator results closely match the execution on the real GPU cluster. In a more realistic scenario, jobs arrive with some batch size distribution
between the minimum and maximum batch sizes. 
Our remaining experiments shall assume jobs with random batch sizes.

\fxfatal{as of now I only see results for category 3. Response: Fixed now.}

\subsection{Effect of different job categories}
Figure~\ref{fig:2} illustrates the total number of jobs completed for various job categories (Section~\ref{sec:jobcategories})
and hence job characteristics. We observe that the results of the simulator is quite close to the real cluster.
We also observe that all the plots, except for Category 4, show that the elastic approach is  more
effective as it demonstrates a significant improvement in the number of jobs completed 
in comparison to the baseline.
The number of jobs completed improve by 82\%, 64.4\% and 90\% for Categories 1, 2 and 3 respectively 
(Figure~\ref{fig:2}(a-c)).
The reason, as discussed before, is that with high arrival rate
the baseline is able to schedule only a limited number of jobs as the batch-size is high and thus
the jobs cannot be scaled down to a small number of GPUs. Our elastic scaling algorithm, however,  
can reduce the batch size of the problem itself to fit the minimum number of GPUs
feasible with the minimum batch-size specified.
Thus it minimizes the number of dropped (rejected) jobs.
To further support this argument, we examined the best throughput scaling factors
for category 1 (compute bound) and category 2 (communication bound) jobs
with their minimum batch size. We observed that the best throughput scaling factor
for the category 1 job was 1.3x the best throughput scaling factor for the category 2 job.

For the 4th category (Figure~\ref{fig:2}(d)), results for both the baseline and our method 
match with each other.
This is because in this category, the jobs are non elastic, i.e. the batch size cannot be changed,
and hence our elastic algorithm does no better than the baseline.

\subsection{Effect of arrival patterns}
Next, we analyze the effect of elastic scaling with different job arrival rates.
Figure~\ref{fig:3}(a) and (b) plot the jobs completed for our elastic scaling algorithm
and the baseline algorithm for micro-benchmarks
wherein jobs arrive with random batch-sizes under low and bursty arrival patterns respectively. The bursty job arrival pattern is obtained by having a high job arrival rate for the first one hour followed by a low job arrival rate for the next one hour.
This pattern is repeated over the duration of 4 hours.

For the case of low arrival rate, the number of jobs completed are more by
97\% ($\approx$ 2 $\times$) with the elastic approach in comparison to the baseline. This is because
both the algorithms try to maximize the GPUs used; however, the elastic scaling algorithm
is able to increase the batch size as it scales to larger number of GPUs whereas the
baseline algorithm cannot. As the scaling behaviour is better when there is more
work to be done by every GPU, the scaling algorithm is able to achieve 
better scaling resulting in quicker completion of the jobs.
This better scaling results in reduced drop rate.
To further support this argument, we present the throughput scaling factors for 
a category 1 jobs on 2 GPUs with different batch-size-per-GPU in Table~\ref{table:sf_cat1_2GPU}.
It is clear that the throughput scaling factor increases monotonically as the batch-size-per-GPU is increased.

When the arrival pattern is bursty, the number of jobs completed by the elastic approach
is 119\% ( $\approx$ 2.2 $\times$) more in comparison to the baseline. 
This is due to a mix of two effects that have already been discussed before. 
During periods of low arrival rate, the elastic scaling algorithm
can increase the batch size as it scales to larger number of GPUs to get better scalability.
On the other hand, during periods of high arrival rate, the elastic scaling algorithm is able to
reduce the batch-size in order to fit on fewer number of GPUs, thereby accommodating more jobs. We also see that the simulator closely match the cloud results.

\subsection{Effect of Queuing}
We next study the effectiveness of our algorithm with queuing; no jobs are dropped (rejected) in this case.
We consider a longer bursty job arrival pattern wherein the jobs arrive for 720 mins (12 hours). 
Every new job that arrives belongs to one of the four
categories with equal probability.
The job lengths (time) for categories 1, 2, 3 and 4 when scheduled on a single GPU
are 16, 21, 41 and 27 mins respectively; these are determined from practical settings
involving incremental training.
This provides a good mix of jobs from different categories with different run times. 
The bursty job arrival pattern is obtained by having a very high job arrival rate for
the first two hours followed by a very low job arrival rate for the next two hours.
This pattern is repeated over the duration of 12 hours.

\begin{table*}
\centering
\begin{tabular}{ |c|cc|cc|cc| } 
 \hline
{} & \multicolumn{2}{c|}{\bf SJS Efficiency (\%)} & \multicolumn{2}{c|}{\bf Job Drop Ratio (\%)} & \multicolumn{2}{c|}{\bf Avg. Job Completion Time (mins) } \\ 
\hline
{} & {Actual} & {Simulator} & {Actual} & {Simulator} & {Actual} & {Simulator} \\ \hline 
 Elastic-withdrop & 82.02 & 80.46 & 13.59 & 10.09 & 24.97 & 22.98\\  \hline
 Baseline-withdrop & 51.31  & 50.49 & 42.4 & 39.57 &  34.12 & 28.10\\ \hline
 Elastic-nodrop & 89.53 & 86.03 & -  & - & 33.79 & 27.75 \\ \hline
 Baseline-nodrop & 42.87 & 41.94 & - & - & 351.02 & 320.32 \\
 \hline
\end{tabular} 
\caption{Performance Metrics for 40 GPUs with and without Job Drops (Random-BS baseline)}
\label{eval:table_util_40GPU}
\end{table*}

\begin{figure}[htb]
\centering
	\subfigure[t][Allow Dropping of Jobs]
	{
		\includegraphics[width=0.4\textwidth, keepaspectratio=true]{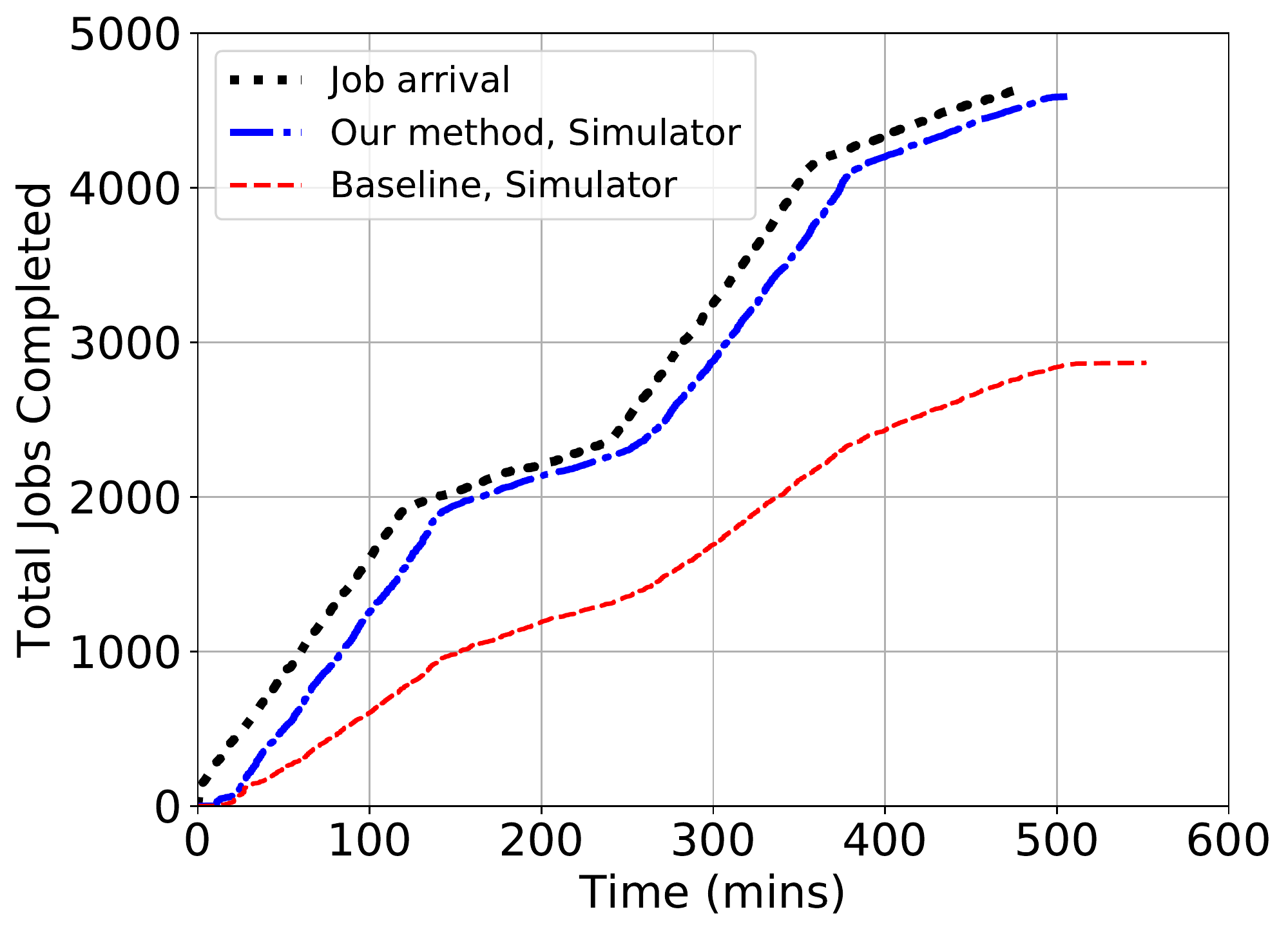}
		\label{plot_400GPU_job_completed_randomBS_bursty}
	}
    \\
	\subfigure[t][No Dropping of Jobs (Queuing of jobs) ]
	{
		\includegraphics[width=0.4\textwidth, keepaspectratio=true]{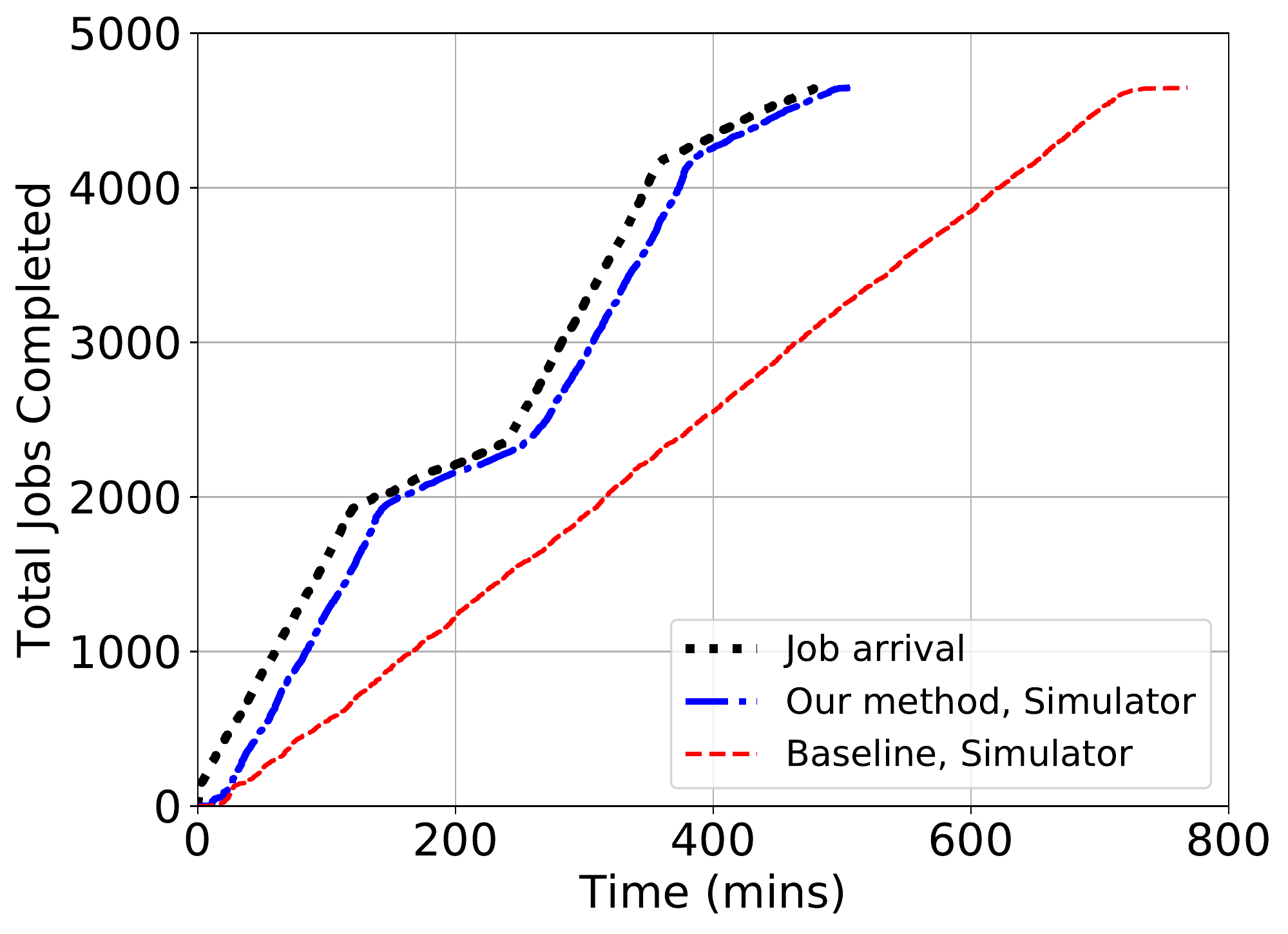}
		\label{plot_400GPU_job_completed_randomBS_bursty_queue}
	}%
\caption{
Simulation results for jobs Completed with bursty job arrival on 400 GPUs (Random-BS baseline)
}%
\label{res:fig_400GPU}%
\label{fig:400}
\vspace{-1mm}
\end{figure}

\begin{table}[htb]
\centering
\scalebox{0.9}{
\begin{tabular}{|c|c|c|c|} 
 \hline
{} & {\bf SJS} & {\bf Job Drop } & {\bf Avg. Job  } \\ 
{} & {\bf Efficiency} & {\bf Ratio} & {\bf Compl. Time} \\  \hline
 Elastic-withdrop & 81.00\% & 1.23\% & 22.83\\  \hline
 Baseline-withdrop & 46.64\% & 38.28\% &  27.84\\ \hline
 Elastic-nodrop & 81.53\% & - & 22.96 \\ \hline
 Baseline-nodrop & 43.10\% & - & 166.82 \\
 \hline
\end{tabular} }
\caption{Performance Metrics for 400 GPUs with and without Job Drops (Random-BS baseline)}
\label{eval:table_util_400GPU}
\label{table:400}
\vspace{-1mm}
\end{table}

The results for this experiment without and with queuing are presented in 
Figure~\ref{res:fig_40GPU_bursty}(a) and (b) respectively. 
We observe that the simulated results are consistent with the actual results on the cloud.  
Our elastic approach performs much better than the baseline;
the number of jobs completed is 50\% more and almost matching the job arrival rate with queuing.
As explained before, 
during periods of low arrival rate, the elastic scaling algorithm can increase the batch size 
as it scales to larger number of GPUs to get better scalability and during periods of high arrival rate, 
the elastic scaling algorithm is able to reduce the batch-size in order to fit on fewer number of GPUs,
thereby accommodating more jobs.

Analyzing the baseline more carefully, 
we see that without queuing the baseline curve tends to flatten (reduction in slope)
during periods of low arrival indicating that it is able to schedule all the jobs.
On the other hand, with queuing the baseline curve remains straight (same slope)
throughout indicating that it is scheduling jobs that have accumulated
in the queue during the period of high arrival. 
In contrast, the elastic scheduling curve flattens in both settings, with and without queuing,
indicating that it is able to service all the queued up jobs.
Thus, the baseline ends up taking considerably long to complete all the jobs.
This is further reflected in the average job completion time for the jobs
presented in Table~\ref{eval:table_util_40GPU}. The average job completion time increases by $\sim$35\%
(24.97 mins to 33.79 mins) for the elastic scaling algorithm whereas it increases by
nearly 10X (34.12 mins to 351.02 mins) for the baseline algorithm. 
The average job completion time for the baseline is nearly 10X that of the elastic
scaling algorithm.

\subsection{Validation of Simulator}
We have seen that the simulation run curves follow
the actual run curves very closely for all the experiments presented so far.
Across all the runs presented, the jobs completed for the simulation runs are
within 7\% of the actual runs. 
Further, we compare other metrics between the simulator run and actual run
for the experiment conducted on 40 GPUs (discussed above) in Table~\ref{eval:table_util_40GPU}.
We can see that the simulator error is less than 5\% for Scheduled Job Scaling (SJS) Efficiency
across all the runs. The difference between the dropped (rejected) jobs ratio
is less than 4\%.
The difference in the average job completion time 
are lower than the actual run by $\sim$10\% for two experiments and $\sim$17\% for two other experiments.
The differences are attributed to the following reasons:
(1) the actual run needs to restart from a previous checkpoint 
(therefore suffering some work loss) every time it is restarted whereas the simulator does not, and
(2) the actual run suffers other system overheads and OS latencies.

We thus conclude that the simulator provides a very good estimate of the 
number of completed jobs, SJS efficiency and dropped job ratio. 
It also provides a reasonable estimate of the average job completion time.

\subsection{Large scale simulation}
We finally evaluate the efficacy of our algorithm on a larger setup comprising of 400 GPUs
using the simulator.
In this setting, the jobs arrive with a bursty arrival pattern 
for 480 mins (8 hours). Every new job that arrives belongs to one of the four
categories with equal probability.
The job lengths (time) for categories 1, 2, 3 and 4 when scheduled on a single GPU
are 16, 21, 41 and 27 mins respectively. 
The bursty job arrival pattern is obtained by having a very high job arrival rate for
the first two hours followed by a very low job arrival rate for the next two hours.
This pattern is repeated over the duration of 8 hours.

The results for this experiment without and with queuing are presented in 
Figure~\ref{fig:400}(a) and (b) respectively. 
The observations are quite similar to those of the 40 GPU experiments.
Our elastic approach performs much better than the baseline;
the number of jobs completed is 60\% more than the baseline and 
almost matching the job arrival rate with queuing.
As in the case of 40 GPUs,
we see that without queuing the baseline curve tends to flatten
during periods of low arrival, whereas with queuing, 
the baseline curve remains straight; 
this indicates that the baseline cannot schedule all the jobs in the queue
during periods of low arrival.
In contrast, the elastic scheduling curve flattens in both settings, with and without queuing,
indicating that it is able to service all the queued up jobs.

The average job completion time for the jobs is
presented in Table~\ref{table:400}. We see that the average job completion time is almost the 
same for the elastic scaling algorithm whereas it increases by
nearly 6X (27.84 mins to 166.82 mins) for the baseline algorithm. 
The average job completion time for the baseline is nearly 6X that of the elastic
scaling algorithm. Note that though the simulator has some error in comparison to the actual runs 
as discussed previously, the error is very small compared to the order of magnitude
difference in comparison between the elastic scaling algorithm and baseline algorithm
discussed here.

\todo[inline, color=green]{Emphasize that performance benefits are minimal only when the cluster remains heavily saturated, AND ALL users choose to use the minimum acceptable batch size - an unrealistic combination in practice.}

\todo[inline]{JSA Accuracy: We have tested and compared the JSA predictions with actual runtimes of jobs in a considerable number of experiments and the predictions are within 3\% of actual wallclock times. Moreover, the close agreement of the simulator and cluster runs for the same workloads presented in the paper validate the accuracy of the JSA. The topic of the paper not being the JSA, we chose not to delve into JSA testing details. In principle, any JSA that can correctly order a set of jobs based on their scalability suffices for the autoscaler.}

\section{Discussion}

\subsection{Choosing Good Min-BS and Max-BS}~\label{app:minmax}
This is a continuation of the discussion in Section~\ref{sec:minmax} and \ref{sec:runtimeestimation}.
One question is whether the specification of Min- and Max-BS to the autoscaler is an undue burden
on data scientists. We believe that it is not.

It is well known in deep learning that each neural network model achieves maximum accuracy when
trained on a given set of batch sizes~\cite{smith-1}. Even when DL training is performed on bare metal servers,
without any elastic scaling or deep learning
platforms, data scientists (need to be) are aware of this range because they need to be ready to train on new data
and the number of GPUs available to them changes either due to budgetary issues or competition from their
colleagues in the same organization. Our work simply leverages this knowledge for effective elastic scaling.

Furthermore, as discussed in Section~\ref{sec:designoverview}, the range of Min- and Max-BS ($b_{min}...b_{max}$)
directly impacts the execution time of the training job. By giving $b_{min}...b_{max}$ as input to the autoscaler,
the data scientists implicitly agrees that they are willing to wait as long as it takes for the job to execute, even in the
worst case that the cluster is extremely overloaded and the autoscaler is only able to allocate the
GPUs corresponding to $b_{min}$. If the data scientist has time constraints, our \JSA\ can also be used to estimate the 
time taken to execute the job for various batch sizes. This is a straightforward extension of the following
iteration time estimation equation from Section~\ref{sec:runtimeestimation}:

\[
t^{\mathit{iter}}_j(b,k)=t^{\mathit{proc}}_j(\lceil \frac{b}{k} \rceil)+t^{\mathit{comm}}(p_j,k)
\]

Thus, the data scientist can then eliminate unsuitable batch sizes according to their time constraints. The 
effort required on the part of the data scientist is minimal considering the performance benefits.

\subsection{Choosing $\triangle$}
\label{app:delta}

This is a continuation of the discussion in Sections~\ref{sec:optimizer} and \ref{sec:autoscaler}.
As discussed earlier in Section~\ref{sec:autoscaler}, to prevent thrashing, the autoscaler
only calls the optimizer every $\triangle$ time interval. If $\triangle$ is too small, then
jobs that are autoscaled will not have made progress or taken checkpoints beyond the last time 
they were halted. If $\triangle$ is too large, this can lead to bad user experience, as some jobs
may spend a lot of time waiting in a queue before being scheduled or rejected. 10-15 minutes seems to
be a good value for $\triangle$ based on our experience. There is also the option of using a hybrid strategy, e.g., 
wait 10 minutes or until 5\% of the jobs in the cluster terminate.

\section{Related Work}~\label{sec:related}

The bulk of existing research (e.g.,~\cite{bauer-chameleon, herbst-tompecs-metric, Ilyushkin-TOMPECS-autoscaler-eval, patrick,google-wavelets,anshul-autoscale, elmem, elasticrmi, elasticity-icdcs16,milind}) on elasticity has focused on cluster elasticity 
Mechanisms for cluster elasticity do not 
adapt the workload to best use a fixed (or semi-fixed) set of resources.
Recent research has also proposed programming elasticity directly into the application,
i.e., making the application self-adaptive by monitoring its environment~\cite{elasticrmi, elasticity-icdcs16, patrick, milind}.
Though programmable elasticity potentially applies to both cluster and job elasticity, aforementioned 
research~\cite{elasticrmi, elasticity-icdcs16, patrick, milind} is limited to cluster elasticity.
Recent research on using burstable instances~\cite{bhuvan-burstable, bhuvan-spot-burstable} is a form of
cluster elasticity, focusing on a combination of vertical and horizontal scaling.

MapReduce~\cite{abhishek-mapred-elasticity, hadoopmr} and Apache Spark~\cite{spark}
were two of the first data analytics platforms to explore job elasticity in a limited manner.
However, these platforms do not handle DL workloads natively, and do not change hyperparameters during elastic scaling. 



As mentioned earlier, Project Philly/Fiddle~\cite{projectphilly, fiddle-atc} is a DL platform internal to Microsoft 
intended for use in datacenter 
and primarily private cloud environments. It enhances YARN's~\cite{yarn} scheduler
to support locality-aware scheduling and gang scheduling. However, elastic scaling is not 
supported. The Hemingway tool~\cite{pan:17} guides the selection of appropriate algorithms 
and cluster size for distributed training jobs by using predictive models 
for job runtime and convergence~\cite{venkataraman:16}. However, these models
do not account for resource contention among jobs in a cloud environment. 
The SLAQ framework~\cite{zhang:17} explores
quality-runtime tradeoffs across multiple jobs to maximize system-wide 
quality by adjusting resource allocations of all
running jobs. SLAQ is a cluster scheduling system for ML training jobs 
that aims to maximize the overall job quality. 
The intuition behind SLAQ is that in the context of 
approximate ML training, more resources should be allocated to jobs that 
have the most potential for quality improvement. 

\fxfatal{we can use more related references, just googled a bit, pasting some of them in the slack channel which are quite relevant.. I have added  \cite{tiresias} \cite{litz} to the bibliography}

\emph{Litz}\cite{litz}, similarly, is one of the first examples of adaptive scaling of resources applied to machine learning (ML) jobs. The Litz environment supports API driven programming of ML constructs, specifically elastic addition/deletion of parameter servers and worker shards. Its benefit is that it can handle elasticity of individual jobs seamlessly (without restarting), but the elasticity decisions are still manual, and there is no concept of a cluster manager/autoscaler that decides the elastic scaling to improve cluster level throughput. 
\emph{Tiresias} \cite{tiresias} proposes a GPU cluster manager for distributed DL jobs which efficiently schedules and places DL jobs to reduce their job completion times (JCT). Their scheduling system assign priorities to jobs in order to reduce their JCT. It doesn't change the batch size for reducing JCT. 
\emph{OASiS}\cite{oasis} describes scheduling in a machine learning cluster that forecasts resource availability and uses that in conjunction with the utility of arriving jobs to arrive at a resource configuration for a particular job. In addition, an admission policy for the job is also prescribed in order to prioritize most useful jobs.


Gandiva~\cite{gandiva} 
exploits intra-job predictability of mini-batch iterations
to time-slice GPUs efficiently across multiple jobs in order to improve cluster efficiency.
It also dynamically migrates a communication intensive 
job to preserve affinities between learners. Gandiva also supports limited
elastic scaling by allowing a learner in a DL training job to occupy all
GPUs on a machine. However, it does not alter hyperparameters during this
limited elastic scaling.

Next, we discuss commercial offerings.
Many public cloud vendors offer DL platforms with some support for elasticity.
Examples include IBM Watson Machine Learning~\cite{ibm-wml}, 
Amazon AWS SageMaker~\cite{amazon-sagemaker}, Google Cloud AI Platform~\cite{google-cloud}, 
and Microsoft Azure~\cite{microsoft-azure}. 
Amazon AWS SageMaker~\cite{amazon-sagemaker} is a fully managed machine learning service,
supporting creation, training and deployment of machine and deep learning models.
On the training side, which is the topic of this paper, Sagemaker only supports
manual elastic scaling. It is up to the data scientist to choose the type and
number of AWS instances for his training job, and determine appropriate hyperparameters.
In the case of Google Cloud AI Platform, training 
either happens via pre-configured VM images optimized for GPUs and TPUs, or 
through Kubeflow~\cite{kubeflow} on Kubernetes. In the case of VM images, elastic
scaling is manual and reactive (train, test, train...). In the case of Kubeflow,
elastic scaling is still user-driven but is \emph{programmable} as part of the 
Kubeflow pipeline. But, neither Kubeflow nor Kubernetes automatically determines the amount of resources to allocate or the hyperparameters corresponding to the allocated resources. 
The elastic scaling capabilities of Microsoft Azure~\cite{microsoft-azure}
with respect to DL training jobs are similar to AWS Sagemaker (other capabilities are 
different, of course).

\section{Conclusions and Future Work}

In this paper, we have demonstrated that to effectively scale deep learning
jobs, we have to determine whether they are compute or communication bound and
leverage the fact that they can be executed with a range of
batch sizes in addition to considering other factors like cluster utilization.
We have demonstrated that the exploration of a range of batch sizes can be
formulated as an optimization problem, which also considers the scaling efficiency
of individual DL jobs when run on multiple resources. We have shown that this problem
admits a dynamic programming based solution which makes it efficient (for real-time use from
autoscalers) and scalable (to typical DL clusters with hundreds of GPUs). 
We have also demonstrated that our elastic scaling techniques have significant performance
benefits in most scenarios (\emph{up to} $2\times$ increase in job completions and $10\times$ better job completion time). 


We are currently in the process of extending this work in two directions -- (i) to support 
job priorities during elastic scaling and (ii) to support spot pricing models to
optimize the cost of job execution (both from the angle of reducing cost to data scientists
and increasing revenue to cloud providers who offer DL-as-a-Service).


\bibliographystyle{plain}
\bibliography{middleware2019}

\cleardoublepage



\end{document}